\documentclass[]{raa}
\usepackage[noblocks]{authblk}
\usepackage{cases}            
\usepackage{graphicx,times}
\usepackage{url}
\usepackage{pdflscape}
\usepackage{longtable}
\usepackage{multirow}

\begin{document}

\title{Comparison between TeV and non-TeV BL Lac Objects}

\volnopage{ {\bf 201X} Vol.\ {\bf XX} No. {\bf XX}, 000--000}
\setcounter{page}{1}

\author{C. Lin
  \inst{1,2}
\and J. H. Fan
  \inst{1,2}
  }

\institute{Center for Astrophysics, Guangzhou University,  Guangzhou 510006, China; {\it fjh@gzhu.edu.cn}
    \and
    Astron. Sci. and Tech. Research Lab. of Dept. of Edu. of Guangdong Province, China \vs \no \\
    {\small Received [year] [month] [day]; accepted [year] [month] [day]}
}

\abstract{
BL Lacertae objects (BL Lacs) is the dominant population of TeV emitting blazars. In this work, we investigate whether there is any special observational properties for TeV sources. To do so, we will compare the observational properties of TeV detected BL Lacs (TeV BLs) and non-TeV detected BL Lac objects (non-TeV BLs). From the 3rd $Fermi$/LAT catalog (3FGL), we can get   662 BL Lacs, out of which, 47 are TeV BLs and 615 are non-TeV BLs. Their  multi-wavelength flux densities ($F_{\rm R}$, $F_{\rm O}$, $F_{\rm X}$, $F_{\gamma}$),  photon spectral indexes ($\alpha_{\rm X}^{\rm ph}$, $\alpha_{\gamma}^{\rm ph}$), and effective spectral indexes ($\alpha_{\rm RO}$ and $\alpha_{\rm OX}$) are compiled from the available literatures.  Then the luminosities ($\log\,{\nu}L_{\rm R}$, $\log\,{\nu}L_{\rm O}$, $\log\,{\nu}L_{\rm X}$, $\log\,{\nu}L_{\gamma}$) are calculated. From comparisons, we found that TeV BLs are different from low-synchrotron-peaked BLs (LSP) and  intermediate-synchrotron-peaked BLs (ISP), but TeV BLs show similar properties as high-synchrotron-peaked BLs (HSP). Therefore, we concentrated on comparison between TeV HSP BLs and non-TeV HSP BLs.  Analysis results suggest that TeV HSP BLs and non-TeV HSP BLs show some differences in their $\alpha_{\rm RO}$ and $\alpha_{\gamma}^{\rm ph}$, while their other properties are quite similar.
}

\authorrunning{C. Lin \& J. H. Fan}            
\titlerunning{Comparison between TeV and non-TeV BLs }  
   \maketitle


%
%
\section{Introduction}  
\label{sect:intro}
Blazars are a subclass of active galactic nuclei (AGNs). They have high and variable polarization, large and rapid variation, superluminal motions, and high energetic GeV (even TeV) $\gamma$-ray emissions, etc (Wills et al. 1992; Zhang \& Fan 2008; Gupta et al. 2008; Romero et al. 2002; Abdo et al. 2010a; Basteri et al. 2011; Fan et al. 2013a, 2014; Ackermann et al. 2015). Blazars can be divided into two subclasses, namely, flat spectrum radio quasars (FSRQs) and BL Lacertae objects (BL Lacs). BL Lacertae objects show no (or very weak) emission line features while FSRQs  display strong emission lines. However, BL Lacs and FSRQs show quite similar continuum emission properties.  BL Lacs can be divided  into radio selected BL Lacertae objects (RBLs) and  X-ray selected BL Lacertae objects (XBLs) from surveys or
 low synchrotron peaked (LSP, $\nu_{\rm peak}^{s} <10^{14}$ Hz),
 intermediate synchrotron peaked  (ISP, $10^{14}$ Hz $<\nu_{\rm peak}^{s} < 10^{15}$ Hz), and
 high synchrotron peaked (HSP, $\nu_{\rm peak}^{s} > 10^{15}$ Hz) BL Lacs from the term by
  Abdo et al. (2010b) (also see
  Fan et al. 2014, 2016;
  Ackermann et al. 2015).
  In 2015, we set different boundaries: LSP ($\nu_{\rm peak}^{s} <10^{14}$ Hz), ISP ($10^{14}$ Hz $<\nu_{\rm peak}^{s} < 10^{16}$ Hz), and HSP ($\nu_{\rm peak}^{s} > 10^{16}$ Hz) (Fan et al. 2015).
  Thanks to the work of the Energetic Gamma Ray Experiment Telescope (EGRET), the $\gamma$-ray astronomy has made great strides forward. As the second generation of $\gamma$-ray detector,  $Fermi$/LAT satellite was launched on June 11, 2008, which detected many blazars at $\gamma$-ray energies (
   Abdo et al. 2010a;
   Nolan et al. 2012;
   Lott et al. 2014; Acero et al. 2015;
   Ackermann et al. 2015). The 3rd  $Fermi$ Large Area Telescope source catalog (3FGL) includes 3033 sources in the 100 MeV--300 GeV ranges, such a large sample of sources gives us a nice opportunity to analyze the nature for  $\gamma$-ray emissions in blazars.

 High energetic emissions as high as TeV are also detected from some blazars,  and  most of them have also been detected by $Fermi$/LAT (Ackermann et al. 2015).
 From TeVCat \footnote{\url{http://tevcat.uchicago.edu/}}, we found that there are 176 sources detected in TeV energy range until March, 2016. These sources include  TeV catalogs which called ¡°Default Catalog¡± and ¡°Newly Announced¡±. The energy threshold for the TeVCat sources is not uniform, but the energy is typically greater than 100 GeV (Acero et al. 2015).  The known extragalactic TeV sources are mainly BL Lacs, but we want to know what kind of BL Lacs to be TeV emitters? So we compared the known TeV BLs with BL Lacs detected in $Fermi$/LAT.

In this work, we compiled BL Lacs from 3FGL and then compared  their observational properties between TeV BL Lacs and  non-TeV BL Lacs, and  try to see whether there is any difference between them. In section 2, we will give a sample, in section 3, we show some results, and in section 4, we will give some discussions and  conclusions.

\section{Sample}

Based on the 3FGL (Acero et al. 2015), the third catalog of AGNs detected by the $Fermi$-LAT (3LAC) is presented (Ackermann et al. 2015). 3LAC not only presents the  $\gamma$-ray data of AGNs detected by $Fermi$--LAT during the first 4 years, but also collects  the fluxes at different bands (radio, optical, and X-ray) and some other data. From 3LAC, we can get 662 Fermi BL Lacs, and their redshift, SED classifications (based on the synchrotron peak frequency), radio flux ($F_{\rm R}$) at 1.4 GHz,
 SDSS V band magnitude ($m_{\rm V}$),
X-ray flux ($F_{\rm X}$) at 0.1--2.4 keV, $\gamma$-ray flux ($F_{\gamma}$) at 1--100 GeV, $\gamma$-ray power-law photon index ($\alpha_{\gamma}^{\rm ph}$) and the effective spectral indexes ($\alpha_{\rm RO}$ and $\alpha_{\rm OX}$) for  BL Lacs are  from the 3LAC Website version\footnote{\url{  http://www.asdc.asi.it/fermi3lac/}}. For some sources in 3LAC, if there are not available data in 3LAC Website, we have looked for them in the NASA/IPAC Extragalactic Database (NED)\footnote{\url{http://ned.ipac.caltech.edu/ }}.
 For optical data, if V band magnitude is not available in 3LAC and NED, we use R band magnitude from NED and $\alpha_O = 1.0$ to estimate it. For the  662 Fermi BL Lacs, 332 BL Lacs have available SDSS V band magnitude from 3LAC, and 24/60 BL Lacs have available V/R-band magnitude from NED.
  For X-ray data, if X-ray flux is not available in 3LAC,
  we compiled the data from the BZCAT version 5.0.0 \footnote{\url{http://www.asdc.asi.it/bzcat}} (Massaro et al. 2015) and NED (April, 2015), and the corresponding X-ray photon indexes ($\alpha_{\rm X}^{\rm ph}$) are from the corresponding references. {

 Combining the 662 Fermi BL Lacs and the TeV sources listed in Table 13 of  a paper by Ackermann et al. (2015), we can get a sample of 47  BL Lacs with both  TeV and GeV emissions, which are listed in Table 1.  Since the TeV sample of Ackermann et al. (2015) is from TeVCat, so the TeV sources in this work are the ones detected in  the range of $E \geq 100$ GeV. For the remaining 615 Fermi BL Lacs that have no TeV emissions, we did  not list their data in the present work.

  \begin{landscape}
  \begin{table} \small
  \begin{center}
{\Large Table 1. TeV sample of the BL Lacs.}\\
\end{center}
\begin{tabular}{cccccccccccccccccc} \hline
3FGL Name&z&SED&$\log\nu^{s}_{p}$&$\alpha_{\gamma}^{ph}$&$F_{X}$&Ref$_1$&$\alpha_{RO}$&$\alpha_{OX}$&TeV &$\alpha_{X}^{ph}$&Ref$_2$&$\log\nu L_{R}$ &$\log\nu L_{O}$ &$\log\nu L_{X}$ &$\log\nu L_{\gamma}$&Other name\\\hline
(1)&(2)&(3)&(4)&(5)&(6)&(7)&(8)&(9)&(10)&(11)&(12)&(13)&(14)&(15)&(16)&(17)\\\hline
3FGL J0013.9-1853	&0.094 	&HSP	&16.76 	&1.94 	&12.60 	&LAC	&-0.01 	&1.58 	&T	&	&	&39.89 	&	&43.79 	&42.89 	 &RBS 0030	&\\
3FGL J0033.6-1921	&0.610 	&HSP	&16.13 	&1.71 	&14.90 	&LAC	&0.20 	&1.03 	&T	&	&	&41.38 	&	&45.78 	&45.71 	 &KUV 00311-1938	&\\
3FGL J0035.9+5949	&	&HSP	&17.12 	&1.90 	&31.80 	&LAC	&	&	&T	&1.42 	&G02	&42.03 	&45.84 	&45.90 	&45.48 	 &1ES 0033+595	&\\
3FGL J0136.5+3905	&	&HSP	&16.20 	&1.70 	&23.30 	&LAC	&0.29 	&1.04 	&T	&2.16 	&B97b	&41.64 	&	&45.71 	 &45.67 	&B3 0133+388	 &\\
3FGL J0152.6+0148	&0.080 	&HSP	&15.46 	&1.89 	&6.07 	&LAC	&0.20 	&1.45 	&T	&2.48 	&B97b	&40.07 	&44.78 	 &43.27 	&43.24 	&PMN J0152+0146	&\\
3FGL J0222.6+4301	&0.444 	&HSP	&15.09 	&1.94 	&6.39 	&LAC	&0.40 	&1.49 	&T	&2.20 	&G02	&43.18 	&46.23 	 &45.10 	&46.31 	&3C 66A	 &\\
3FGL J0232.8+2016	&0.139 	&HSP	&15.48 	&2.03 	&15.70 	&LAC	&0.23 	&1.22 	&T	&1.99 	&D05	&40.69 	&45.00 	 &44.37 	&43.53 	&1ES 0229+200	&\\
3FGL J0303.4-2407	&0.260 	&HSP	&15.43 	&1.92 	&10.20 	&LAC	&0.45 	&1.20 	&T	&2.68 	&F12	&42.18 	&	&44.57 	 &45.24 	&PKS 0301-243	 &\\
3FGL J0319.8+1847	&0.190 	&HSP	&16.99 	&1.57 	&27.00 	&LAC	&0.30 	&0.80 	&T	&1.50 	&G02	&40.41 	&44.59 	 &44.98 	&43.85 	&RBS 0413	 &\\
3FGL J0349.2-1158	&0.185 	&HSP	&18.29 	&1.73 	&30.90 	&LAC	&0.23 	&0.83 	&T	&2.03 	&D05	&40.42 	&44.54 	 &44.93 	&43.61 	&1ES 0347-121	&\\
3FGL J0416.8+0104	&0.287 	&HSP	&16.64 	&1.75 	&73.20 	&LAC	&0.33 	&0.82 	&T	&2.80 	&R00	&41.50 	&45.27 	 &45.48 	&44.33 	&1ES 0414+009	&\\
3FGL J0449.4-4350	&0.205 	&HSP	&15.67 	&1.85 	&14.30 	&LAC	&0.38 	&1.19 	&T	&	&	&41.70 	&	&44.60 	&45.24 	 &PKS 0447-439	&\\
3FGL J0508.0+6736	&0.340 	&HSP	&17.75 	&1.52 	&35.90 	&LAC	&0.30 	&0.82 	&T	&2.31 	&G02	&41.00 	&	&45.53 	 &44.93 	&1ES 0502+675	 &\\
3FGL J0521.7+2113	&0.108 	&ISP	&14.38 	&1.92 	&6.02 	&LAC	&	&	&T	&1.21 	&B97b	&41.27 	&	&43.82 	&44.54 	 &TXS 0518+211	&\\
3FGL J0550.6-3217	&0.069 	&	&	&1.61 	&51.20 	&LAC	&0.11 	&1.46 	&T	&2.28 	&D05	&40.68 	&44.16 	&44.14 	 &42.58 	&PKS 0548-322	 &\\
3FGL J0648.8+1516	&0.179 	&HSP	&15.92 	&1.83 	&38.10 	&LAC	&	&	&T	&	&	&40.81 	&	&44.89 	&44.25 	&RX J0648.7+1516	&\\
3FGL J0650.7+2503	&0.203 	&HSP	&16.42 	&1.72 	&42.30 	&LAC	&0.28 	&1.02 	&T	&2.47 	&F12	&41.09 	&	&45.02 	 &44.52 	&1ES 0647+250	 &\\
3FGL J0710.3+5908	&0.125 	&HSP	&16.99 	&1.66 	&32.50 	&LAC	&0.19 	&1.14 	&T	&2.15 	&F12	&40.88 	&	&44.54 	 &43.51 	&1H 0658+595	 &\\
3FGL J0721.9+7120	&0.127 	&LSP	&13.99 	&2.04 	&4.91 	&LAC	&0.43 	&1.50 	&T	&2.10 	&R00	&41.55 	&45.38 	 &43.75 	&45.15 	&S5 0716+71	 &\\
3FGL J0809.8+5218	&0.138 	&HSP	&15.86 	&1.88 	&17.80 	&LAC	&0.33 	&1.21 	&T	&3.00 	&R00	&41.02 	&45.05 	 &44.00 	&44.52 	&1ES 0806+524	&\\
3FGL J0847.1+1134	&0.199 	&HSP	&15.95 	&1.74 	&23.80 	&LAC	&0.36 	&0.80 	&T	&2.50 	&B97b	&40.61 	&44.62 	 &44.74 	&43.87 	&RX J0847.1+1133	&\\
3FGL J1010.2-3120	&0.143 	&HSP	&16.31 	&1.58 	&28.30 	&LAC	&0.14 	&1.31 	&T	&	&	&40.67 	&	&44.54 	&43.71 	 &1RXS J101015.9-311909	 &\\
3FGL J1015.0+4925	&0.212 	&HSP	&15.63 	&1.83 	&19.80 	&LAC	&0.39 	&1.16 	&T	&2.48 	&F12	&41.73 	&45.33 	 &44.73 	&45.12 	&1H 1013+498	&\\
3FGL J1103.5-2329	&0.186 	&HSP	&17.19 	&1.64 	&50.90 	&LAC	&0.32 	&0.88 	&T	&2.25 	&G02	&41.11 	&	&45.09 	 &43.68 	&1ES 1101-232	 &\\
3FGL J1104.4+3812	&0.031 	&HSP	&17.07 	&1.77 	&678.00 	&LAC	&0.29 	&0.92 	&T	&2.82 	&F12	&40.33 	&44.60 	 &44.28 	&43.96 	&Mkn 421	&\\
3FGL J1136.6+7009	&0.045 	&HSP	&15.77 	&1.82 	&56.70 	&LAC	&0.02 	&1.70 	&T	&2.20 	&R00	&40.28 	&43.94 	 &43.83 	&42.94 	&Mkn 180	 &\\
3FGL J1217.8+3007	&0.130 	&HSP	&15.26 	&1.97 	&86.40 	&LAC	&0.42 	&0.94 	&T	&2.47 	&B00	&41.47 	&44.99 	 &44.89 	&44.60 	&1ES 1215+303	&\\
3FGL J1221.3+3010	&0.182 	&HSP	&16.66 	&1.66 	&31.60 	&LAC	&0.30 	&0.98 	&T	&2.10 	&B00	&40.86 	&44.99 	 &44.91 	&44.58 	&PG 1218+304	&\\
3FGL J1221.4+2814	&0.103 	&ISP	&14.42 	&2.10 	&2.29 	&LAC	&0.47 	&1.58 	&T	&2.10 	&R00	&41.37 	&44.89 	 &43.23 	&44.24 	&W Comae	 &\\
3FGL J1224.5+2436	&0.218 	&HSP	&15.39 	&1.89 	&2.75 	&LAC	&0.29 	&1.27 	&T	&2.22 	&B00	&40.59 	&44.88 	 &43.99 	&44.14 	&MS 1221.8+2452	&\\\hline
\end{tabular}
\end{table}
\end{landscape}
\begin{landscape}
\begin{table} \small
  \begin{center}
{\Large Table 1. Continue.}\\
\end{center}
\begin{tabular}{cccccccccccccccccc} \hline
3FGL Name&z&SED&$\log\nu^{s}_{p}$&$\alpha^{ph}_{\gamma}$&$F_{X}$&Ref$_1$&$\alpha_{RO}$&$\alpha_{OX}$&TeV &$\alpha^{ph}_{X}$&Ref$_2$&$\log\nu L_{R}$ &$\log\nu L_{O}$ &$\log\nu L_{X}$ &$\log\nu L_{\gamma}$&Other name\\\hline
(1)&(2)&(3)&(4)&(5)&(6)&(7)&(8)&(9)&(10)&(11)&(12)&(13)&(14)&(15)&(16)&(17)\\\hline
3FGL J1427.0+2347	&	&HSP	&15.34 	&1.82 	&6.94 	&LAC	&0.32 	&1.51 	&T	&2.54 	&B00	&42.49 	&46.57 	&45.08 	 &46.16 	&PKS 1424+240	 &\\
3FGL J1428.5+4240	&0.129 	&HSP	&17.28 	&1.57 	&52.50 	&LAC	&0.28 	&0.87 	&T	&1.92 	&F12	&40.47 	&44.69 	 &44.85 	&43.53 	&H 1426+428	 &\\
3FGL J1442.8+1200	&0.163 	&HSP	&16.35 	&1.80 	&13.80 	&LAC	&0.37 	&0.93 	&T	&2.20 	&F12	&40.75 	&44.60 	 &44.41 	&43.74 	&1ES 1440+122	&\\
3FGL J1517.6-2422	&0.048 	&ISP	&14.19 	&2.11 	&2.92 	&LAC	&0.20 	&2.18 	&T	&2.36 	&F12	&41.14 	&44.38 	 &42.54 	&43.64 	&AP Librae	 &\\
3FGL J1555.7+1111	&0.360 	&HSP	&15.47 	&1.68 	&38.60 	&LAC	&0.31 	&1.30 	&T	&2.50 	&R00	&42.12 	&46.48 	 &45.57 	&45.84 	&PG 1553+113	&\\
3FGL J1653.9+3945	&0.034 	&HSP	&16.12 	&1.72 	&65.10 	&LAC	&0.40 	&1.20 	&T	&2.36 	&F12	&40.72 	&44.57 	 &43.57 	&43.54 	&Mkn 501	 &\\
3FGL J1725.0+1152	&0.018 	&HSP	&16.01 	&1.89 	&32.00 	&LAC	&0.29 	&1.16 	&T	&2.65 	&F12	&39.05 	&43.62 	 &42.56 	&42.57 	&1H 1720+117	&\\
3FGL J1728.3+5013	&0.055 	&HSP	&16.00 	&1.96 	&39.60 	&LAC	&0.20 	&1.30 	&T	&2.39 	&F12	&40.25 	&43.77 	 &43.78 	&43.06 	&I Zw 187	 &\\
3FGL J1743.9+1934	&0.084 	&HSP	&15.76 	&1.78 	&11.80 	&LAC	&0.08 	&1.88 	&T	&1.98 	&B97b	&40.81 	&	&43.78 	 &43.18 	&S3 1741+19	 &\\
3FGL J2000.0+6509	&0.047 	&HSP	&16.86 	&1.88 	&114.00 	&LAC	&0.07 	&1.46 	&T	&2.68 	&F12	&40.20 	&45.05 	 &43.96 	&43.64 	&1ES 1959+650	&\\
3FGL J2001.1+4352	&	&HSP	&15.21 	&1.97 	&1.00 	&BZC	&	&	&T	&	&	&41.87 	&	&44.30 	&45.99 	&MG4 J200112+4352	&\\
3FGL J2009.3-4849	&0.071 	&HSP	&16.29 	&1.77 	&80.80 	&LAC	&0.19 	&1.56 	&T	&2.05 	&G02	&41.28 	&45.07 	 &44.44 	&43.77 	&PKS 2005-489	&\\
3FGL J2158.8-3013	&0.116 	&HSP	&15.97 	&1.83 	&572.00 	&LAC	&0.20 	&1.01 	&T	&2.57 	&G02	&41.30 	&45.76 	 &45.56 	&45.02 	&PKS 2155-304	&\\
3FGL J2202.7+4217	&0.069 	&LSP	&13.61 	&2.25 	&7.42 	&LAC	&0.43 	&1.70 	&T	&2.63 	&G02	&41.93 	&44.64 	 &43.15 	&44.48 	&BL Lacertae	&\\
3FGL J2250.1+3825	&0.119 	&	&	&1.91 	&7.93 	&LAC	&0.20 	&1.51 	&T	&2.51 	&B97b	&40.65 	&	&43.75 	&43.77 	 &B3 2247+381	&\\
3FGL J2347.0+5142	&0.044 	&HSP	&15.87 	&1.78 	&29.70 	&LAC	&	&	&T	&2.13 	&F14	&40.15 	&43.95 	&43.55 	 &43.18 	&1ES 2344+514	 &\\
3FGL J2359.3-3038	&0.165 	&HSP	&17.52 	&2.02 	&65.00 	&LAC	&0.35 	&0.65 	&T	&1.82 	&F12	&40.73 	&44.10 	 &45.19 	&43.83 	&H 2356-309	 &\\\hline
\end{tabular}
\end{table}
{\footnotesize Note to the Table:
 Col. (1) gives a 3FGL name,
 Col. (2)  redshift,
 Col. (3) SED classification,
 Col. (4) synchrotron peak frequency ($\log\nu^{s}_{p}$) in the unit of Hz from 3LAC, the $\log\nu^{s}_{p}$ are already corrected by redshift in 3LAC.
 Col. (5) $\gamma$-ray photon index,
 Col. (6) and (7)   X-ray flux in units of $10^{-12}$ erg/cm$^2$/s at 0.1--2.4 keV and the corresponding references,
 Col. (8) and (9)   effective spectral indexes ($\alpha_{RO}$ and $\alpha_{OX}$),
 Col. (10) `T" stands for TeV sources,
 Col. (11) and (12)  X-ray photon index and the corresponding references,
 Col. (13), (14), (15) and (16) give  radio, optical, X-ray at 1 keV, and $\gamma$-ray (at 2 GeV) luminosities ($\log\nu L_{\nu}$) in units of erg/s,
 Col. (17) other names.
 Here A09: Ajello et al. (2009); B00: Brinkmann et al. (2000); B97a: Brinkmann et al. (1997a); B97b: Brinkmann et al. (1997b); BZC: Massaro et al. (2015); D05: Donato et al. (2005); F12: Fan et al. (2012); F13: Fan et al. (2013a); G09: Green et al. (2009); LAC: Ackermann et al. (2015); L96: Lamer et al. (1996); L99: Laurent et al. (1999); NED: the NASA/IPAC Extragalactic Database (\url{http://ned.ipac.caltech.edu/}); R00: Reich et al. (2000)\\
}
\end{landscape}

\section{Results}

In this work, luminosity is calculated using $\nu L_{\nu}=4\pi d_{\rm L}^2\nu F_{\nu}$, where $d_L$ is the luminosity distance. The Cosmology Calculator I\footnote{\url{http://www.astro.ucla.edu/~wright/CosmoCalc.html}} from NED is used to calculate the luminosity distance (Wright 2006),
 here we adopt $H_0 = 73$ km/s/Mpc, $\Omega_{\rm M} = 0.27$, $\Omega_{\rm vac} = 0.73$.
All the fluxes are K-corrected, the flux in the source rest frame is $F_{\nu}^{\rm res} = F_{\nu}^{\rm obs}(1+z)^{\alpha-1}$, $\alpha$ ($F_{\nu} \propto \nu^{-\alpha}$) is the energy spectral index (Kapanadze 2013).
 $\alpha_{\rm R}$ = 0.0 and $\alpha_{\rm O}$ = 1.0 are adopted for radio and optical bands,
 $\alpha_{\rm X}$ = $\alpha_{\rm X}^{\rm ph} - 1$,  $\alpha_{\gamma}$ = $\alpha_{\gamma}^{\rm ph} - 1$.    Most of the X-ray spectral indexes are given for 0.1--2.4 keV. If there is no spectral index information in the 0.1--2.4 keV band, and there is a spectral index in hard X-ray band, then we use the spectral index in the hard X-ray band instead. If  redshift and X-ray photon index are unkown, then averaged values, $<z>$ = 0.463 and $<\alpha_{\rm X}^{\rm ph}>\, =\, 2.35$ are adopted.  For optical V-band luminosity calculation, V band magnitude ($m_{\rm V}$) is transferred into flux density ($F_{\rm V}$)  using $m_{\rm V} \,= \, 16.40-2.5\log\, F_{\rm V}$, where $F_{\rm V}$ is the flux density in  units of mJy (Kapanadze 2013). All the V band magnitudes are corrected by Galactic Extinction from NED.

For the 662 BL Lacs in 3LAC, there are 286 HSP, 185 ISP, 168 LSP, and 23 for unknown SED type. Out of the 662 BL Lacs, there are  47 TeV BL Lacs (including 40 HSP, 3 ISP, 2 LSP, 2 unknown SED type). For the TeV BLs and non-TeV BLs, we made some comparisons for  z, $\alpha_{\rm X}^{\rm ph}$, $\alpha_{\gamma}^{\rm ph}$, $\log{\nu}L_{\rm R}$, $\log{\nu}L_{\rm O}$, $\log{\nu}L_{\rm X}$, $\log{\nu}L_{\gamma}$, $\alpha_{\rm RO}$, and $\alpha_{\rm OX}$ as follows.

\subsection{Averaged Values}

For the whole sample, the redshift is in a range of 0.002 $\, \leq \, z \,\leq 2.471$;
 X-ray and $\gamma$-ray photon spectral indexes are 1.03 $\, \leq \, \alpha_{\rm X}^{\rm ph} \,\leq 4.28$ and 1.26 $\, \leq \, \alpha_{\gamma}^{\rm ph} \,\leq 2.81$; radio, optical, X-ray and $\gamma$-ray luminosities are in the ranges:
 36.30 erg/s$\, \leq \, \log{\nu}L_{\rm R} \,\leq 44.10$ erg/s,
 40.29 erg/s$\, \leq \, \log{\nu}L_{\rm O} \,\leq 47.03$ erg/s,
 39.45 erg/s$\, \leq \, \log{\nu}L_{\rm X} \,\leq 46.45$ erg/s, and
 39.24 erg/s$\, \leq \, \log{\nu}L_{\gamma} \,\leq 47.33$ erg/s;  effective spectral indexes satisfy
  $-$0.13 $\, \leq \, \alpha_{\rm RO} \,\leq 0.96$, and 0.43 $\, \leq \, \alpha_{\rm OX} \,\leq 2.52$.
The corresponding averaged values  are listed in Table 2.
And the corresponding Kolmogorov-Smirnov (K-S) test results are listed in Table 3, in which,
  Col. (1) gives two tested samples,
Col. (2) tested parameter,
Col. (3) number of two samples,
Col. (4) averaged values and standard deviation,
Col. (5) K-S statistics $d_{\rm max}$,
Col. (6) two-tailed significance probability $p$.
The sample with ``$\ast$" is only for the sources with available redshift, $p$ is the probability for the two distributions to come from the same distribution. The corresponding histograms and cumulative distributions are shown in Figs. \ref{Lin-2015-TeV-z-K-S}--\ref{Lin-2015-TeV-alpha_OX-K-S}.

From Tables 2 and 3, and the corresponding Figures \ref{Lin-2015-TeV-z-K-S}--\ref{Lin-2015-TeV-alpha_OX-K-S}, we can see that, for redshift, TeV BLs are clearly different from non-TeV BLs, HSP BLs and ISP+LSP BLs with probabilities for the corresponding two groups to come from the same distribution being $p\, < \, 10^{-5}$, suggesting that the redshift of TeV BLs is lower than those of the  rest groups;
  For X-ray photon index, TeV BLs are not different from  non-TeV BLs, HSP BLs or ISP+LSP  BLs;
  For $\gamma$-ray photon index and radio luminosity, TeV BLs are clearly different from non-TeV BLs and ISP+LSP BLs, but not different from HSP BLs; For optical luminosity, TeV BLs are marginally  different from non-TeV  and ISP+LSP BLs, but not different from HSP BLs;
  For  X-ray luminosity and effective optical-X-ray spectral index, there is  no much difference between TeV and non-TeV BLs or HSP BLs, but TeV BLs are different from ISP+LSP  BLs;
  For $\gamma$-ray luminosity and effective radio-optical spectral index, TeV BLs are clearly different from non-TeV BLs and ISP+LSP BLs, but only marginally different from HSP BLs.

\subsection{Correlations between $\gamma$-ray and other bands}

Now we adopted a linear regression analysis to  fluxes and luminosities to investigate the correlationship between $\gamma$-ray and other bands. For luminosity-luminosity correlations, we only considered the sources with available redshift, and obtained

 log $ \nu L_{\gamma}$ = ( 1.052 $\pm$ 0.099 ) log $\nu\,L_{R}$ + ( 1.087 $\pm$ 4.041 ) for 36 TeV HSP BLs with a correlation coefficient $r =  0.877$ and a chance probability of $p \, = \, 2.31 \times 10^{-12}$, and

 log $ \nu L_{\gamma}$ = ( 0.981 $\pm$ 0.047 ) log $\nu L_{R}$ + ( 3.963 $\pm$ 1.913 ) for 157 non-TeV HSP BLs with  $r =  0.861$ and $p   = \, 2.85 \times 10^{-47}$;

 log $ \nu L_{\gamma}$ = ( 1.152 $\pm$ 0.117 ) log $\nu L_{O}$ $-$ ( 7.611 $\pm$ 5.242 ) for 25 TeV HSP BLs with  $r =  0.899$ and $p \, = \, 1.02 \times 10^{-9}$, and

 log $ \nu L_{\gamma}$ = ( 1.069 $\pm$ 0.059 ) log $\nu L_{O}$ $-$ ( 3.926 $\pm$ 2.659 ) for 95 non-TeV HSP BLs with  $r =  0.883$ and  $p \, = \, 2.89 \times 10^{-32}$;

 log $ \nu L_{\gamma}$ = ( 0.869 $\pm$ 0.149 ) log $\nu L_{X}$ + ( 5.375 $\pm$ 6.633 ) for 36 TeV HSP BLs with  $r =  0.707$ and  $p \, = \, 1.40 \times 10^{-6}$, and

 log $ \nu L_{\gamma}$ = ( 0.756 $\pm$ 0.048 ) log $\nu L_{X}$ + ( 10.638 $\pm$ 2.143 ) for 151 non-TeV BLs with  $r =  0.788$ and  $p \, = \, 2.95 \times 10^{-33}$.

  The flux densities in unit of mJy are calculated in radio at 1.4 GHz, optical at V-band, X-ray at 1 keV, and $\gamma$-ray band at 2 GeV.
  For flux-flux correlations, we have

 log $ F_{\gamma}$ = ( 0.752 $\pm$ 0.126 ) log $F_{R}$ $-$ ( 10.794 $\pm$ 0.280 ) for 40 TeV HSP BLs with a correlation $r =  0.696$ and a chance probability of $p \, = \, 6.00 \times 10^{-7}$, and
 log $ F_{\gamma}$ = ( 0.299 $\pm$ 0.045 ) log $F_{R}$ $-$ ( 10.250 $\pm$ 0.076 ) for 246 non-TeV HSP BLs with  $r =  0.389$ and  $p \, = \, 2.66 \times 10^{-10}$;

 log $ F_{\gamma}$ = ( 0.798 $\pm$ 0.113 ) log $F_{O}$ $-$ ( 9.590 $\pm$ 0.109 ) for 27 TeV HSP BLs with  $r =  0.769$ and  $p \, = \, 2.76 \times 10^{-6} $, and
 log $ F_{\gamma}$ = ( 0.318 $\pm$ 0.066 ) log $F_{O}$ $-$ ( 9.780 $\pm$ 0.030 ) for 131 non-TeV HSP BLs with  $r =  0.392$ and  $p \, = \, 3.73 \times 10^{-6}$;

 log $ F_{\gamma}$ = ( 0.073 $\pm$ 0.184 ) log $F_{X}$ $-$ ( 8.982 $\pm$ 0.480 ) for 40 TeV HSP BLs with  $r =  0.064$ and  $p \, = \, 69.3 \%$, and
 log $ F_{\gamma}$ = ( 0.054 $\pm$ 0.039 ) log $F_{X}$ $-$ ( 9.585 $\pm$ 0.136 ) for 237 non-TeV BLs with  $r =  0.089$ and  $p \, = \, 17.4 \%$.

 log $ F_{O}$ = ( 0.744 $\pm$ 0.147 ) log $F_{R}$ $-$ ( 1.076 $\pm$ 0.340 ) for 27 TeV HSP BLs with  $r =  0.711$ and  of $p \, = \, 3.19 \times 10^{-5}$, and
 log $ F_{O}$ = ( 0.557 $\pm$ 0.071 ) log $F_{R}$ $-$ ( 1.047 $\pm$ 0.116) for 131 non-TeV BLs with  $r =  0.567$ and  $p \, = \, 1.58 \times 10^{-12}$.

All the results are listed in Table 4 and shown in Figs. \ref{Lin-2015-TeV-L-L-corr}--\ref{Lin-2015-TeV-logF-logF-corr}.

\begin{landscape}
\begin{table}
\begin{center}
{\Large Table 2. Averaged values of Fermi BL Lacs.}
\end{center}
\begin{tabular}{l|c|c|c|c|c|c|c|c|c}
        \hline
         & & All & TeV & non-TeV & HSP & ISP+LSP & TeV HSP & non-TeV HSP & Figure\\
         \hline
         (1) & (2) & (3) & (4) & (5) & (6) & (7) & (8) & (9) & (10) \\
         \hline
        \multirow{2}{1cm}{Redshift}
                & Average & 0.463 $\pm$ 0.418 & 0.157 $\pm$ 0.116 & 0.500 $\pm$ 0.426 & 0.336 $\pm$ 0.322 & 0.596 $\pm$ 0.467 & 0.170 $\pm$ 0.123 & 0.347 $\pm$ 0.341 & \multirow{2}{1cm}{Fig. \ref{Lin-2015-TeV-z-K-S}}\\
                & Number & 403 & 43 & 360 & 193 & 197 & 36 & 157\\
                \hline
        \multirow{2}{1cm}{$\alpha^{\rm ph}_{\rm X}$}
                & Average & 2.352 $\pm$ 0.534 & 2.273 $\pm$ 0.369 & 2.371 $\pm$ 0.567 & 2.336 $\pm$ 0.447 & 2.337 $\pm$ 0.644 & 2.295 $\pm$ 0.350 & 2.353 $\pm$ 0.483 & \multirow{2}{1cm}{Fig. \ref{Lin-2015-TeV-alphaX-K-S}}\\
                & Number & 206 & 41 & 165 & 116 & 85 & 34 & 82\\
                \hline
         \multirow{2}{1cm}{$\alpha^{\rm ph}_{\gamma}$}
                & Average & 2.023 $\pm$ 0.246 & 1.827 $\pm$ 0.156 & 2.038 $\pm$ 0.246 & 1.875 $\pm$ 0.200 & 2.143 $\pm$ 0.206 & 1.798 $\pm$ 0.129 & 1.888 $\pm$ 0.207 & \multirow{2}{1cm}{Fig. \ref{Lin-2015-TeV-alpha_G-K-S}}\\
                & Number & 662 & 47 & 615 & 286 & 353 & 40 & 246\\
                \hline
         \multirow{2}{1cm}{$\log{\nu}L_{\rm R}$}
                & Average & 41.599 $\pm$ 0.912 & 41.028 $\pm$ 0.750 & 41.643 $\pm$ 0.909 & 41.119 $\pm$ 0.753 & 42.024 $\pm$ 0.828 & 40.993 $\pm$ 0.789 & 41.139 $\pm$ 0.746 & \multirow{2}{1cm}{Fig. \ref{Lin-2015-TeV-L_R-K-S}}\\
                & Number & 662 & 47 & 615 & 286 & 353 & 40 & 246\\
                \hline
         \multirow{2}{1cm}{$\log{\nu}L_{\rm R}^\ast$}
                & Average &  &  &  &  &  & 40.881 $\pm$ 0.743 & 40.002 $\pm$ 0.850 & \multirow{2}{1cm}{Fig. \ref{Lin-2015-TeV-L_R-K-S}}\\
                & Number &  &  &  &  &  & 36 & 157\\
                \hline
         \multirow{2}{1cm}{$\log{\nu}L_{\rm O}$}
                & Average & 45.213 $\pm$ 0.735 & 44.885 $\pm$ 0.724 & 45.240 $\pm$ 0.730 & 45.083 $\pm$ 0.725 & 45.316 $\pm$ 0.735 & 44.922 $\pm$ 0.763 & 45.116 $\pm$ 0.715 & \multirow{2}{1cm}{Fig. \ref{Lin-2015-TeV-L_O-K-S}}\\
                & Number & 416 & 32 & 384 & 158 & 248 & 27 & 131\\
                \hline
         \multirow{2}{1cm}{$\log{\nu}L_{\rm O}^\ast$}
                & Average &  &  &  &  &  & 44.819 $\pm$ 0.686 & 45.042 $\pm$ 0.793 & \multirow{2}{1cm}{Fig. \ref{Lin-2015-TeV-L_O-K-S}}\\
                & Number &  &  &  &  &  & 25 & 95\\
                \hline
         \multirow{2}{1cm}{$\log{\nu}L_{\rm X}$}
                & Average & 44.325 $\pm$ 0.865 & 44.435 $\pm$ 0.821 & 44.314 $\pm$ 0.869 & 44.529 $\pm$ 0.876 & 44.065 $\pm$ 0.791 & 44.602 $\pm$ 0.748 & 44.517 $\pm$ 0.897 & \multirow{2}{1cm}{Fig. \ref{Lin-2015-TeV-L_X-K-S}}\\
                & Number & 530 & 47 & 483 & 277 & 235 & 40 & 237\\
                \hline
         \multirow{2}{1cm}{$\log{\nu}L_{\rm X}^\ast$}
                & Average &  &  &  &  &  & 44.530 $\pm$ 0.725 & 44.341 $\pm$ 0.997 & \multirow{2}{1cm}{Fig. \ref{Lin-2015-TeV-L_X-K-S}}\\
                & Number &  &  &  &  &  & 36 & 151\\
                \hline
         \multirow{2}{1cm}{$\log{\nu}L_{\gamma}$}
                & Average & 44.712 $\pm$ 0.906 & 44.232 $\pm$ 0.970 & 44.749 $\pm$ 0.892 & 44.400 $\pm$ 0.881 & 44.988 $\pm$ 0.840 & 44.262 $\pm$ 0.999 & 44.423 $\pm$ 0.860 & \multirow{2}{1cm}{Fig. \ref{Lin-2015-TeV-L_G-K-S}}\\
                & Number & 662 & 47 & 615 & 286 & 353 & 40 & 246\\
                \hline
         \multirow{2}{1cm}{$\log{\nu}L_{\gamma}^\ast$}
                & Average &  &  &  &  &  & 44.088 $\pm$ 0.891 & 44.190 $\pm$ 0.969 & \multirow{2}{1cm}{Fig. \ref{Lin-2015-TeV-L_G-K-S}}\\
                & Number &  &  &  &  &  & 36 & 157\\
                \hline
         \multirow{2}{1cm}{$\alpha_{\rm RO}$}
                & Average & 0.425 $\pm$ 0.161 & 0.275 $\pm$ 0.116 & 0.436 $\pm$ 0.159 & 0.328 $\pm$ 0.100 & 0.511 $\pm$ 0.156 & 0.269 $\pm$ 0.111 & 0.337 $\pm$ 0.096 & \multirow{2}{1cm}{Fig. \ref{Lin-2015-TeV-alpha_RO-K-S}}\\
                & Number & 616 & 42 & 574 & 272 & 326 & 36 & 236\\
                \hline
         \multirow{2}{1cm}{$\alpha_{\rm OX}$}
                & Average & 1.275 $\pm$ 0.309 & 1.227 $\pm$ 0.333 & 1.280 $\pm$ 0.307 & 1.154 $\pm$ 0.293 & 1.426 $\pm$ 0.255 & 1.156 $\pm$ 0.289 & 1.154 $\pm$ 0.295 & \multirow{2}{1cm}{Fig. \ref{Lin-2015-TeV-alpha_OX-K-S}}\\
                & Number & 519 & 42 & 477 & 268 & 236 & 36 & 232\\
        \hline
        \end{tabular}
        \end{table}
        {\footnotesize Note to the Table:
Col. (1) gives parameter,
Col. (2) averaged value row or number row of the sample,
Col. (3) all sample,
Col. (4) TeV BLs,
Col. (5) non-TeV BLs,
Col. (6) HSP BLs,
Col. (7) ISP+LSP BLs,
Col. (8) TeV HSP BLs,
Col. (9) non-TeV HSP BLs,
Col. (10) the corresponding Figure number.
For luminosities of HSP BLs, the samples with available redshift are marked by ``$\ast$". The luminosities are in unit of erg/s.
}
        \end{landscape}


\newpage
  \begin{center}
{\Large Table 3. The results of the two samples K-S test.}
\begin{tabular}{lccccc} \hline
samples & parameter & N & averaged & $d_{\rm max}$ & $p$ \\
\hline
(1)&(2)&(3)&(4)&(5)&(6)\\
\hline
TeV / non-TeV&&43 / 360&0.157$\pm$0.116 / 0.500$\pm$0.426&0.613&$<10^{-13}$\\
TeV / HSP&z&43 / 193&0.157$\pm$0.116 / 0.336$\pm$0.322&0.436&$<10^{-5}$\\
TeV / ISP+LSP&&43 / 197&0.157$\pm$0.116 / 0.596$\pm$0.467&0.662&$<10^{-14}$\\
TeV HSP / non-TeV HSP&&36 / 157&0.170$\pm$0.123 / 0.374$\pm$0.341&0.502&$<10^{-6}$\\
\hline
TeV / non-TeV&&41 / 165&2.273$\pm$0.369 / 2.371$\pm$0.567&0.187&14.82\%\\
TeV / HSP&$\alpha^{ph}_{\rm X}$&41 / 116&2.273$\pm$0.369 / 2.336$\pm$0.447&0.097&87.17\%\\
TeV / ISP+LSP&&41 / 85&2.273$\pm$0.369 / 2.377$\pm$0.644&0.233&7.06\%\\
TeV HSP / non-TeV HSP&&34 / 82&2.295$\pm$0.350 / 2.353$\pm$0.483&0.131&70.56\%\\
\hline
TeV / non-TeV&&47 / 615&1.827$\pm$0.156 / 2.038$\pm$0.246&0.471&$<10^{-8}$\\
TeV / HSP&$\alpha^{ph}_{\gamma}$&47 / 286&1.827$\pm$0.156 / 1.875$\pm$0.200&0.137&32.68\%\\
TeV / ISP+LSP&&47 / 353&1.827$\pm$0.156 / 2.143$\pm$0.206&0.691&$<10^{-19}$\\
TeV HSP / non-TeV HSP&&40 / 246&1.798$\pm$0.129 / 1.888$\pm$0.207&0.243&2.08\%\\
\hline
TeV / non-TeV&&47 / 615&41.028$\pm$0.750 / 41.643$\pm$0.909&0.342&$<10^{-4}$\\
TeV / HSP&&47 / 286&41.028$\pm$0.750 / 41.119$\pm$0.753&0.161&20.19\%\\
TeV / ISP+LSP&$\log{\nu}L_{\rm R}$&47 / 353&41.028$\pm$0.750 / 42.024$\pm$0.828&0.528&$<10^{-10}$\\
TeV HSP / non-TeV HSP&&40 / 246&40.993$\pm$0.789 / 41.139$\pm$0.746&0.233&3.67\%\\
TeV HSP$^\ast$ / non-TeV HSP$^\ast$&&36 / 157&40.881$\pm$0.743 / 40.002$\pm$0.850&0.191&19.96\%\\
\hline
TeV / non-TeV&&32 / 384&44.885$\pm$0.724 / 45.240$\pm$0.730&0.302&0.64\%\\
TeV / HSP&&32 / 158&44.885$\pm$0.724 / 45.083$\pm$0.725&0.247&5.79\%\\
TeV / ISP+LSP&$\log{\nu}L_{\rm O}$&32 / 248&44.885$\pm$0.724 / 45.316$\pm$0.735&0.364&0.06\%\\
TeV HSP / non-TeV HSP&&27 / 131&44.922$\pm$0.763 / 45.116$\pm$0.715&0.269&5.85\%\\
TeV HSP$^\ast$ / non-TeV HSP$^\ast$&&25 / 95& 44.819$\pm$0.686 / 45.042$\pm$0.793&0.257&11.00\%\\
\hline
TeV / non-TeV&&47 / 483&44.435$\pm$0.821 / 44.314$\pm$0.869&0.106&65.78\%\\
TeV / HSP&&47 / 277&44.435$\pm$0.821 / 44.529$\pm$0.876&0.105&69.09\%\\
TeV / ISP+LSP&$\log{\nu}L_{\rm X}$&47 / 235&44.435$\pm$0.821 / 44.065$\pm$0.791&0.260&0.78\%\\
TeV HSP / non-TeV HSP&&40 / 237&44.602$\pm$0.748 / 44.517$\pm$0.897&0.093&87.92\%\\
TeV HSP$^\ast$ / non-TeV HSP$^\ast$&&36 / 151&44.530$\pm$0.725 / 44.341$\pm$0.997&0.163&36.32\%\\
\hline
TeV / non-TeV&&47 / 615&44.232$\pm$0.970 / 44.749$\pm$0.892&0.359&$<10^{-4}$\\
TeV / HSP&&47 / 286&44.232$\pm$0.970 / 44.400$\pm$0.881&0.241&1.38\%\\
TeV / HSP($\log\nu^{s}_{p}\geq16$)&$\log{\nu}L_{\gamma}$&47 / 121&44.232$\pm$0.970 / 44.292$\pm$0.681&0.217&6.43\%\\
TeV / ISP+LSP&&47 / 353&44.232$\pm$0.970 / 44.988$\pm$0.840&0.474&$<10^{-8}$\\
TeV HSP / non-TeV HSP&&40 / 246&44.262$\pm$0.999 / 44.423$\pm$0.860&0.293&0.37\%\\
TeV HSP$^\ast$ / non-TeV HSP$^\ast$&&36 / 157&44.088$\pm$0.891 / 44.190$\pm$0.969&0.231&6.97\%\\
\hline
TeV / non-TeV&&42 / 574&0.275$\pm$0.116 / 0.436$\pm$0.159&0.455&$<10^{-7}$\\
TeV / HSP&&42 / 272&0.275$\pm$0.116 / 0.328$\pm$0.100&0.252&1.55\%\\
TeV / HSP($\log\nu^{s}_{p}\geq16$)&$\alpha_{\rm RO}$&42 / 115&0.275$\pm$0.116 / 0.326$\pm$0.093&0.268&1.80\%\\
TeV / ISP+LSP&&42 / 326&0.275$\pm$0.116 / 0.511$\pm$0.156&0.648&$<10^{-14}$\\
TeV HSP / non-TeV HSP&&36 / 236&0.269$\pm$0.111 / 0.337$\pm$0.096&0.282&1.06\%\\
\hline
TeV / non-TeV&&42 / 477&1.227$\pm$0.333 / 1.280$\pm$0.307&0.152&29.75\%\\
TeV / HSP&$\alpha_{\rm OX}$&42 / 268&1.227$\pm$0.333 / 1.154$\pm$0.293&0.165&24.07\%\\
TeV / ISP+LSP&&42 / 236&1.227$\pm$0.333 / 1.426$\pm$0.255&0.364&$<10^{-4}$\\
TeV HSP / non-TeV HSP&&36 / 232&1.156$\pm$0.289 / 1.154$\pm$0.295&0.081&97.34\%\\
\hline
\end{tabular}
\end{center}

\newpage
\begin{table} \small
\begin{center}
{\Large Table 4. Correlations of Fermi HSP BL Lacs.}
\end{center}
\begin{tabular}{c|c|c|c|c|c|c|c}
        \hline
         $y$ & $x$ & samples & $N$ & $A$ & $B$ & $r$ & $p$\\
         \hline
         \multirow{2}{1.4cm}{$\log \nu\,L_{\gamma}$} & \multirow{2}{1.4cm}{$\log \nu\,L_{\rm R}$}
          & TeV HSP & 36 & 1.052 $\pm$ 0.099 & 1.087 $\pm$ 4.041 & 0.877 & $2.31 \times 10^{-12}$\\
          && non-TeV HSP & 157 & 0.981 $\pm$ 0.047 & 3.963 $\pm$ 1.913 & 0.861 & $2.85 \times 10^{-47}$\\
         \hline
         \multirow{2}{1.4cm}{$\log \nu\,L_{\gamma}$} & \multirow{2}{1.4cm}{$\log \nu\,L_{\rm O}$}
          & TeV HSP & 25 & 1.152 $\pm$ 0.117 & $-7.611 \pm 5.242$ & 0.899 & $1.02 \times 10^{-9}$\\
          && non-TeV HSP & 95 & 1.069 $\pm$ 0.059 & $-3.926 \pm 2.659$ & 0.883 & $2.89 \times 10^{-32}$\\
         \hline
         \multirow{2}{1.4cm}{$\log \nu\,L_{\gamma}$} & \multirow{2}{1.4cm}{$\log \nu\,L_{\rm X}$}
          & TeV HSP & 36 & 0.869 $\pm$ 0.149 & 5.375 $\pm$ 6.633 & 0.707 & $1.40 \times 10^{-6}$\\
          && non-TeV HSP & 151 & 0.756 $\pm$ 0.048 & 10.638 $\pm$ 2.143 & 0.788 & $2.95 \times 10^{-33}$\\
          \hline
          \multirow{2}{1.4cm}{$\log F_{\gamma}$} & \multirow{2}{1.4cm}{$\log F_{\rm R}$}
          & TeV HSP & 40 & 0.752 $\pm$ 0.126 & $-10.794 \pm 0.280$ & 0.696 & $6.00 \times 10^{-7}$\\
          && non-TeV HSP & 246 & 0.299 $\pm$ 0.045 & $-10.250 \pm 0.076$ & 0.389 & $2.66 \times 10^{-10}$\\
          \hline
          \multirow{2}{1.4cm}{$\log F_{\gamma}$} & \multirow{2}{1.4cm}{$\log F_{\rm O}$}
          & TeV HSP & 27 & 0.798 $\pm$ 0.113 & $-9.590 \pm 0.109$ & 0.769 & $2.76 \times 10^{-6}$\\
          && non-TeV HSP & 131 & 0.318 $\pm$ 0.066 & $-9.780 \pm 0.030$ & 0.392 & $3.73 \times 10^{-6}$\\
          \hline
          \multirow{2}{1.4cm}{$\log F_{\gamma}$} & \multirow{2}{1.4cm}{$\log F_{\rm X}$}
          & TeV HSP & 40 & 0.073 $\pm$ 0.184 & $-8.982 \pm 0.480$ & 0.064 & $69.3 \%$\\
          && non-TeV HSP & 237 & 0.054 $\pm$ 0.039 & $-9.585 \pm 0.136$ & 0.089 & $17.4 \%$\\
          \hline
          \multirow{2}{1.4cm}{$\log F_{\rm O}$} & \multirow{2}{1.4cm}{$\log F_{\rm R}$}
          & TeV HSP & 27 & 0.744 $\pm$ 0.147 & $-1.076 \pm 0.340$ & 0.711 & $3.19 \times 10^{-5}$\\
          && non-TeV HSP & 131 & 0.557 $\pm$ 0.071 & $-1.047 \pm 0.116$ & 0.567 & $1.58 \times 10^{-12}$\\
        \hline
        \end{tabular}
        \end{table}
        {\footnotesize Note to the Table:
Col. (1) gives dependent parameter,
Col. (2) independent parameter,
Col. (3) samples,
Col. (4) number of the sample,
Col. (5) slope,
Col. (6) intercept,
Col. (7) correlation coefficient,
Col. (8)  chance probability.}

\section{Discussions and  Conclusions }

In this paper, we compiled the multi-wavelength data ($\alpha_{\rm X}^{\rm ph}$, $F_{\rm R}$, $F_{\rm O}$, $F_{\rm X}$, $F_{\gamma}$, $\alpha_{\gamma}^{\rm ph}$, $\alpha_{\rm RO}$ and $\alpha_{\rm OX}$) for a sample of 662 BL Lacs from the 3LAC and some other references, calculated the luminosity and averaged values for z, $\alpha_{\rm X}^{\rm ph}$, $\alpha_{\gamma}^{\rm ph}$, $\log\nu L_{\rm R}$, $\log\nu L_{\rm O}$, $\log\nu L_{\rm X}$, $\log\nu L_{\gamma}$, $\alpha_{\rm RO}$ and $\alpha_{\rm OX}$ for TeV BLs and the subgroups of BLs, and made some comparisons by using K-S test and correlation analysis. The results are listed in Tables 2--4 and shown in  Figs. \ref{Lin-2015-TeV-z-K-S}--\ref{Lin-2015-TeV-alpha_RO-alpha_OX-alpha_G}.

\subsection{\bf Averaged values}

{\it \underline{TeV BLs and non-TeV BLs:\,\,\,}}  From Tables 2 and 3 and  Figs. \ref{Lin-2015-TeV-z-K-S}--\ref{Lin-2015-TeV-alpha_OX-K-S}, we can see clear difference between TeV BLs and non-TeV BLs in
  $\gamma$-ray photon spectral index ($\alpha_{\gamma}$) with a probability for the two groups to come from the same distribution being $p\,\, < \, 10^{-8}$, that is
    $p\, < \, 10^{-13}$ in redshift ($z$),
  $p\, < \, 10^{-4}$ in radio luminosity ($\log\nu L_{\rm R}$) and  $\gamma$-ray luminosity ($\log\nu L_{\gamma}$), and
  $p\, < \, 10^{-7}$ in effective radio-optical spectral index ($\alpha_{\rm RO}$). But there is no clear difference in X-ray photon spectral index ($\alpha_{\rm X}^{\rm ph}$), optical luminosity ($\log \nu L_{\rm O}$), X-ray luminosity ($\log \nu L_{\rm X}$), or effective optical-X-ray spectral index ($\alpha_{\rm OX}$) between  TeV  BLs and non-TeV BLs. The averaged $\gamma$-ray photon spectral indexes show that TeV BLs have harder spectrum than non-TeV BLs, the steeper spectrum of non-TeV BLs makes the TeV emissions be below the sensitivity of TeV detectors so that they can not be detected. For the known TeV BLs, their  redshift are small, it is possible that the TeV emissions from sources with high redshift maybe absorbed by the cosmic background emissions. Therefore, it is hard to detect TeV emissions from high redshift. For non-TeV BLs, they are strongly beamed in radio and $\gamma$-ray bands, so their radio and $\gamma$-ray luminosities are higher than those for TeV BLs. Also, the non-TeV BLs in this work, are mainly LBLs and IBLs, they show their synchrotron peak frequencies in the range of infrared to optical bands. Their radio emissions are luminous, which also result in the difference in the effective spectral index, $\alpha_{\rm RO}$ with  $\alpha_{\rm RO}$ in non-TeV BLs being larger than that in TeV BLs.

{\it \underline{TeV BLs and LBLs/IBLs:\,\,\,}}  The difference between TeV BLs and LBLs/IBLs is clear in $\alpha_{\gamma}^{\rm ph}$, $z$, $\log\nu L_{\rm R}$, $\log\nu L_{\rm O}$, $\log \nu L_{\rm X}$, $\log\nu L_{\gamma}$, $\alpha_{\rm RO}$ and $\alpha_{\rm OX}$, but that is not clear in $\alpha_{\rm X}^{\rm ph}$ ($p = 7.06\%$). For BL Lacs, the X-rays are from synchrotron emissions for HBLs whose peak emissions are in the range of UV/X regions, and the summation of synchrotron emissions and inverse Compton emissions
 for LBL/IBL (Fan et al. 2012), which result in that there is no clear difference in their X-ray spectral indexes.

{\it \underline{TeV BLs and HSP BLs:\,\,\,}}  There is almost no difference between TeV BLs and HSP BLs in $\alpha_{\gamma}^{\rm ph}$,  $\alpha_{\rm X}^{\rm ph}$, $\log\nu L_{\rm R}$, $\log\nu L_{\rm O}$, $\log\nu L_{\rm X}$, or $\alpha_{\rm OX}$, but there is marginal difference in $\log\nu L_{\gamma}$ with a $p = 1.38\%$ and in $\alpha_{\rm RO}$ with a $p = 1.55\%$.  Therefore, when we only  considered  the TeV HSP BLs and non-TeV HSP BLs, we found clear difference in redshift ($z$ with $ p\,\, < 10^{-6}$ ), marginal difference in $\gamma$-ray photon spectral index ($\alpha_{\gamma}$ with $p = 2.08\%$) and radio to optical spectral index ($\alpha_{\rm RO}$ with $p = 1.06\%$), and no difference in other parameters between TeV HSP BLs and non-TeV HSP BLs. In those different parameters ($z$, $\alpha_{\gamma}$, $\alpha_{\rm RO}$), the averaged values of TeV HSP BLs are lower than those of non-TeV HSP BLs. Does that mean HSP BLs with lower redshift, harder $\alpha_{\gamma}$, and smaller $\alpha_{\rm RO}$ are good candidates for TeV emitters? If so, then we can use $\alpha_{\gamma}$, $\alpha_{\rm RO}$, and redshift ($z$) to predict TeV HSP BL Lacs.

From above analyses, we can see clearly that TeV BL Lacs are different from LBLs/IBLs, but are similar to HSP BL Lacs. If we only take HSP BL Lacs into account then we find that TeV HSP BL Lacs are quite similar to non-TeV HSP BL Lacs except for that there is difference in redshift and  marginal differences in  $\alpha_{\gamma}$ and $\alpha_{\rm RO}$ as shown in the lower panel of Fig. \ref{Lin-2015-TeV-alpha_RO-alpha_OX-alpha_G}, which  shows $\alpha_{\rm RO} \leq 0.45$ and $\alpha_{\gamma} \leq 2.03$ for TeV HSP BL Lacs.

Taking only TeV  and non-TeV HSP BLs into account, we can see that the flux densities for TeV HSP BLs are averagely higher than non-TeV HSP BLs: It is $2.161 \pm 0.536$ (TeV HSP) vs $1.613 \pm 0.458$ (non-TeV HSP) for $\log F_{\rm R}$;
$0.594 \pm 0.581$ (TeV HSP) vs $-0.174 \pm 0.422$ (non-TeV HSP) for $\log F_{\rm O}$;
$-2.564 \pm 0.510$ (TeV HSP) vs $-3.419 \pm 0.589$ (non-TeV HSP) for $\log F_{\rm X}$; and
$-9.169 \pm 0.578$ (TeV HSP) vs $-9.768 \pm 0.352$ (non-TeV HSP) for $\log F_{\gamma}$. The difference can also be seen from  Fig. \ref{Lin-2015-TeV-logF-logF-corr}.

Those differences indicate that TeV HSP BLs tend to have higher fluxes than non-TeV ones. The fluxes of non-TeV HSP BLs maybe very low in TeV band so that they can not be detected, thus more sensitive telescopes are required. As we noted, BL Lacs have rapid and large variations, so the TeV emissions are likely to cycle. Therefore, some TeV sources can not be detected in TeV band sometimes, and a long time TeV monitoring programme can detect more BL Lacs. In addition, the difference of fluxes between TeV HSP BLs and non-TeV HSP BLs could be an other criteria to predict TeV emitter candidates.

\subsection{\bf Correlations}

From luminosity-luminosity correlation analysis of TeV HSP BL Lacs, we can see that there is a close correlation between $\gamma$-ray luminosity and the lower energetic bands. Those correlation coefficients and chance probabilities are:
 $r = 0.877$ and $ p = 2.31 \times 10^{-12}$ for $\rm \log \nu L_{\gamma} \,\,vs\,\, \log \nu L_{\rm R}$,
 $r = 0.899$ and $ p = 1.02 \times 10^{-9}$ for $\rm \log \nu L_{\gamma} \,\,vs\,\,  \log \nu L_{\rm O}$, and
 $r = 0.707$ and $ p = 1.40 \times 10^{-6}$ for $\rm \log \nu L_{\gamma} \,\,vs\,\,  \log \nu L_{\rm X}$,
 see Table 4 and Fig. \ref{Lin-2015-TeV-L-L-corr}.
We also found that the relation slopes of correlations for TeV HSP BLs are similar to those for non-TeV HSP BLs, and the $\rm \log \nu L_{\gamma} \,\,vs\,\, \log \nu L_{\rm R}$ correlation is the strongest. The strong $\gamma$-ray vs radio correlation was also discussed by other authors (
 Dondi \& Ghisellini 1995;
 M\"{u}cke et al. 1997;
 Xie et al. 1997;
 Zhou et al. 1997;
 Fan et al. 1998, 2012, 2015, 2016;
 Cheng et al. 2000;
 Yang \& Fan 2005;
 Giroletti et al. 2010, 2012;
 Pushkarev et al. 2010;
 Linford et al. 2011;
 Yang et al. 2012a,b; 2014;
 Li et al. 2015).

For luminosity-luminosity correlation, it is  known that all luminosities are correlated with redshift ($z$), therefore  luminosity-luminosity correlation maybe caused by the redshift effect
(Kendall \& Stuart, 1979). In this case, one should remove the redshift effect.  To do so, we used the method introduced by
 Padovani  (1992) as did in our previous work (Fan et al. 2013b; 2015).
If variables  $i$ and $j$ are correlated with a third one $k$, then the correlation between $i$ and $j$ should exclude the
$k$ effect. In this sense, for three variables of $i$, $j$ and $k$, if the correlation coefficients of relation
between any two variables of them are expressed as $r_{ij}$,
$r_{ik}$, $r_{jk}$ respectively, and after the correlation
coefficient $r_{ij}$ to be excluded the $k$ effect is expressed  as ${r_{ij,k}} = ({r_{ij}} -
{r_{ik}}{r_{jk}})/\sqrt {(1 - r_{ik}^2)(1 - r_{jk}^2)}$. When the
method is applied to the correlations between any two
luminosities, the correlation coefficients after removing the redshift effect
 are:
 For $\rm \log \nu L_{\gamma} \,\,vs\,\, \log \nu L_{\rm R}$,
 $r_{\gamma R, z}$ = 0.743 with a chance probability $p_{\gamma R,z} = 7.17 \times 10^{-7}$ for TeV HSP BLs, and
 $r_{\gamma R, z}$ = 0.400 with $p_{\gamma R,z} = 4.52 \times 10^{-7}$ for non-TeV HSP BLs;
 For $\rm \log \nu L_{\gamma} \,\,vs\,\, \log \nu L_{\rm O}$,
 $r_{\gamma O, z}$ = 0.811 with $p_{\gamma O,z} = 3.52 \times 10^{-6} $ for TeV HSP BLs, and
 $r_{\gamma O, z}$ = 0.375 with $p_{\gamma O,z} = 3.0\times 10^{-4} $ for non-TeV HSP BLs; and
 for $\rm \log \nu L_{\gamma} \,\,vs\,\, \log \nu L_{\rm X}$,
 $r_{\gamma X, z}$ = 0.255 with $p_{\gamma X,z} = 9.3 \%$ for TeV HSP BLs,   and
 $r_{\gamma X, z}$ = 0.085 with $p_{\gamma X,z} = 15.91 \%$ for non-TeV HSP BLs.
 It is clear that after removing the redshift effect, there is still correlations for $\rm \log \nu L_{\gamma} \,\,vs\,\, \log \nu L_{\rm R}$ and for $\rm \log \nu L_{\gamma} \,\,vs\,\, \log \nu L_{\rm O}$ for TeV HSP  BL Lacs, but there is no correlation for
 $\rm \log \nu L_{\gamma} \,\,vs\,\, \log \nu L_{\rm X}$. The results are consistent with those for flux-flux correlation analysis: namely
 $r\, = \, 0.696$ with $p\,=\, 6.00 \times 10^{-7}$ for $\rm \log F_{\gamma} \,\,vs\,\, \log F_{\rm R}$,
 $r\, = \, 0.769$ with $p\,=\, 2.76 \times 10^{-6}$ for $\rm \log F_{\gamma} \,\,vs\,\, \log F_{\rm O}$, and
 $r\, = \, 0.064$ with $p\,=\, 69.3\%$ for $\rm \log F_{\gamma} \,\,vs\,\, \log F_{\rm X}$.
 For the non-TeV HSP BLs, results show that the correlation coefficients are less  than those in TeV HSP after removing the redshift effect, and  all the correlation coefficients are less than 0.4, which is very different from the correlation before removing the redshift effect. The results are also consistent with those for flux-flux correlation analysis:
 $r\, = \, 0.389$ with $p\,=\, 2.66 \times 10^{-10}$ for $\rm \log F_{\gamma} \,\,vs\,\, \log F_{\rm R}$,
 $r\, = \, 0.392$ with $p\,=\, 3.73 \times 10^{-6}$ for $\rm \log F_{\gamma} \,\,vs\,\, \log F_{\rm O}$, and
 $r\, = \, 0.089$ with $p\,=\, 17.4\%$ for $\rm \log F_{\gamma} \,\,vs\,\, \log F_{\rm X}$.

For flux-flux correlations, both TeV HSP BLs and non-TeV HSP BLs show strong correlations in $\log F_{\gamma}$ vs $\log F_{\rm R}$, $\log F_{\gamma}$ vs $\log F_{\rm O}$, and $\log F_{\rm O}$ vs $\log F_{\rm R}$ with chance probabilities being $p < 10^{-4}$. In those flux-flux correlations, the slopes of TeV HSP BLs are steeper than non-TeV HSP BLs, but those intercepts are very close, which suggest that TeV HSP BLs tend to have higher energy photons (Fig. \ref{Lin-2015-TeV-logF-logF-corr}). However,  for both TeV and non-TeV HSP BLs, no correlation was found between $\log F_{\gamma}$ and $\log F_{\rm X}$.

 Dondi \& Ghisellini (1995)  found the correlations between $\gamma$-ray luminosity and lower energy bands (including radio, optical and X-ray band) of blazars, and the $\gamma$-ray vs radio band correlation still exist even after removing redshift effect. But they did not discuss the correlations of $\gamma$-ray vs optical or $\gamma$-ray vs X-ray since they think that optical and X-ray emissions are contaminated by other emissions. Our results are consistent with theirs. The correlation between X-ray and $\gamma$-ray emissions are obtained by some surveys in a long period of observations or some samples of blazars (Li et al. 2013; Bi et al. 2014; Fraija et al. 2015).
 For example, Fraija et al. (2015) found a strong correlation between the GeV $\gamma$-rays and the optical/hard-X ray emissions for Mrk 421 at the flare in 2013. However, no correlation was found between $\gamma$-ray and the X-ray band for PKS 1510-089 during its high activity period from 2008 to 2009 (Abdo et al. 2010c).
 From our analysis, there is no correlation between $\gamma$-ray and X-ray flux suggesting that the $\gamma$-ray and X-ray emissions are composed of different emission components even though the X-ray emissions in  HSP BL Lac are from synchrotron self-Compton (SSC) process (Fan et al. 2012).

For spectral index correlation, we can see that there is an anti-correlation between $\alpha_{OX}$ and $\alpha_{RO}$ for TeV HSP BL Lacs and non-TeV HSP BL Lacs:
 $\alpha_{\rm OX}$ = $-$( 1.445 $\pm$ 0.373) $\alpha_{\rm RO}$ + ( 1.545 $\pm$ 0.108 ) for 36 TeV HSP BL Lacs with a correlation coefficient $r = - 0.554$ and a chance probability $p \, = \, 4.59 \times 10^{-4}$, and
  $\alpha_{\rm OX}$ = $-$( 1.422 $\pm$ 0.180 ) $\alpha_{\rm RO}$ + ( 1.634 $\pm$ 0.063 ) for 232 non-TeV HSP BL Lacs with a correlation coefficient $r = - 0.462$ and a chance probability $p \, = \, 1.10 \times 10^{-13}$.
 The corresponding plots are shown in the upper panel of Fig. \ref{Lin-2015-TeV-alpha_RO-alpha_OX-alpha_G}.

\subsection{ Mechanism}

 $\gamma$-ray emissions are still an interesting topic for blazars, and the Fermi mission has provided us a good opportunity to re-visit the $\gamma$-ray mechanism by detecting a lot of blazars ( Abdo et al 2010a,
 Nolan et al. 2012,
 Acero et al. 2015;
 Ackermann et al. 2015).   As we discussed in our previous work (Fan et al. 2013a, 2014), the $\gamma$-ray emissions is mainly due to soft photons upscattered by Inverse Compton onto relativistic electrons, or to synchrotron emission/pion decay of secondary particles produced in a proton-induced cascade (PIC) (Mannheim \& Biermann 1992; Mannheim, 1993; Cheng \& Ding 1994). For LBLs, they have low peak synchrotron  emissions with log $\nu_p\,< 14.0 \rm{Hz}$ and their inverse Compton emissions peak at $\rm log \nu_{IC} < 1 GeV$ while for HBLs, they have a synchrotron peak frequency of $\rm log \nu_p > 15 \rm{Hz}$ and their inverse Compton emissions peak at $\rm log \nu_{IC} > 100 GeV$ (Abdo et al. 2009). Therefore, the  emissions in the 1$\sim$100 GeV region  correspond to the  inverse-Compton emission tail, which have a soft spectrum for LBLs, and the emissions in the 1$\sim$100 GeV region  correspond to the  inverse-Compton emissions before reaching the peak emissions and have a flat spectrum for HBLs. That is why TeV HSP BLs have flat spectrum (Fan et al. 2012). In this sense, we think that SSC will be responsible for $\gamma$-ray emissions for HBLs and particularly for TeV HSP BLs.
 Fortunately, the spectral energy distributions of some blazars can be fitted by a one-zone SSC model in some survey (Sambruna et al. 2000; Albert et al. 2007; Aleksi\'c et al. 2015a). Aleksi\'c et al. (2015b) found that the SSC model gives a satisfactory description of the observed multi-wavelength spectral energy distribution for PG 1553+113 during the flare.
 Zhang et al. (2012) compiled the broadband SEDs data for 24 TeV BL Lac objects and found that these SEDs can be explained well with the SSC model.  Abdo et al. (2014) believed the TeV emissions from Mrk 421 are produced by leptonic SSC emissions. After 3 years of observations of Mrk 421, they found that the TeV activity, measured as what is called duty cycle, is consistent with the X-ray activity and therefore favors the SSC emission mechanism. If we use the SSC model to explain the TeV radiation, the HSP BL Lacs, which have high-synchrotron-peaked frequency, will have more probability to produce TeV radiation.  Our results of $\gamma$-ray and radio correlations obtained by luminosity-luminosity  and flux-flux relationships also support an SSC process for $\gamma$-rays.

\subsection{Conclusion}

  In this work, we compiled the radio, optical, X-ray and $\gamma$-ray data for a sample of 662 Fermi BL Lacs (47 are TeV BL Lacs, and 615 are non-TeV BL Lacs) from 3LAC (Ackermann et al. 2015) and other references, calculated the flux density and luminosity, compared the averaged values and investigated luminosity-luminosity and flux-flux  correlations for TeV BL Lacs and subclasses of BL Lacs. Following conclusions have been come to:

  1) TeV BL Lacs are different from LBLs and IBLs in the distributions of $\alpha_{\gamma}^{\rm ph}$, $z$, $\log\nu L_{\rm R}$, $\log\nu L_{\rm O}$, $\log\nu L_{\rm X}$, $\log\nu L_{\gamma}$, $\alpha_{\rm RO}$ and $\alpha_{\rm OX}$, but not from $\alpha_{\rm X}^{\rm ph}$. TeV BL Lacs tend to show similar properties of HSP BL Lacs in $\alpha_{\gamma}^{\rm ph}$,  $\alpha_{\rm X}^{\rm ph}$, $\log\nu L_{\rm R}$, $\log\nu L_{\rm O}$, $\log\nu L_{\rm X}$ and $\alpha_{\rm OX}$, but not in $\log\nu L_{\gamma}$ or $\alpha_{\rm RO}$;

  2) TeV HSP BL Lacs show different distributions of redshift from non-TeV HSP BL Lacs, marginal different distributions in $\alpha_{\rm RO}$ and $\alpha_{\gamma}$, but no difference in other parameters. So, HSP  BL Lacs with low redshift, $\alpha_{\rm RO}$, $\alpha_{\gamma}$, and high fluxes are  good TeV emitter candidates;

  3) There is a significant correlation between $\gamma$-ray and radio bands and between $\gamma$-ray and optical bands, but there is no correlation $\gamma$-ray and X-ray bands for TeV HSP BLs;

  4)  The $\gamma$-ray emissions in HSP BLs are from SSC model.

Acknowledgements: The work is partially supported by the
National Natural Science Foundation of China (NSFC 10633010, NSFC
11173009, U1431112, U1531245), the Innovation Foundation of Guangzhou University (IFGZ),
Guangdong Province Universities and Colleges Pearl River Scholar
Funded Scheme(GDUPS)(2009), Yangcheng Scholar Funded
Scheme(10A027S), and support for Astrophysics  Key Subjects of Guangdong Province and Guangzhou City.
We thank the referee for the comments and suggestions which  improve the manuscript.

 \clearpage
 \begin{figure}[h]
 \centering
 \includegraphics[height=4.50in]{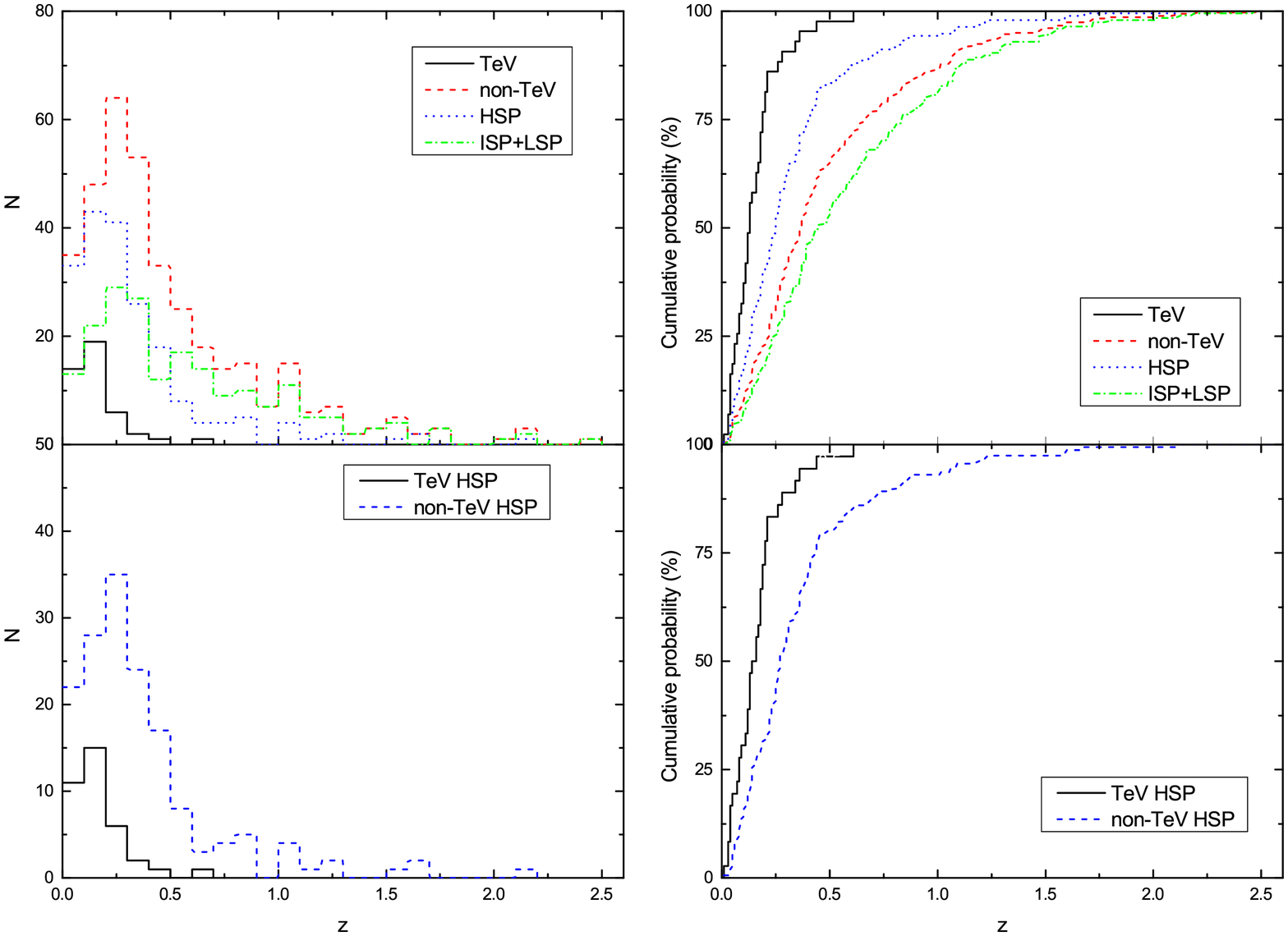}\\
 \caption{The distribution of the redshift (left panel) and the accumulative probability (right panel) for the whole sample (upper sub-panel), and for the HSP BL Lacs (lower sub-panel). In the upper sub-panel, both left and right panel, solid line stands for TeV sources, broken line for non-TeV sources, dotted line for HSP, broken-dotted line for ISP+LSP. In the lower sub-panel, both left and right panel, solid line stands for TeV HSP BL Lacs, broken line for non-TeV HSP BL Lacs.}
 \label{Lin-2015-TeV-z-K-S}
 \end{figure}

 \begin{figure}[h]
\centering
 \includegraphics[height=4.50in]{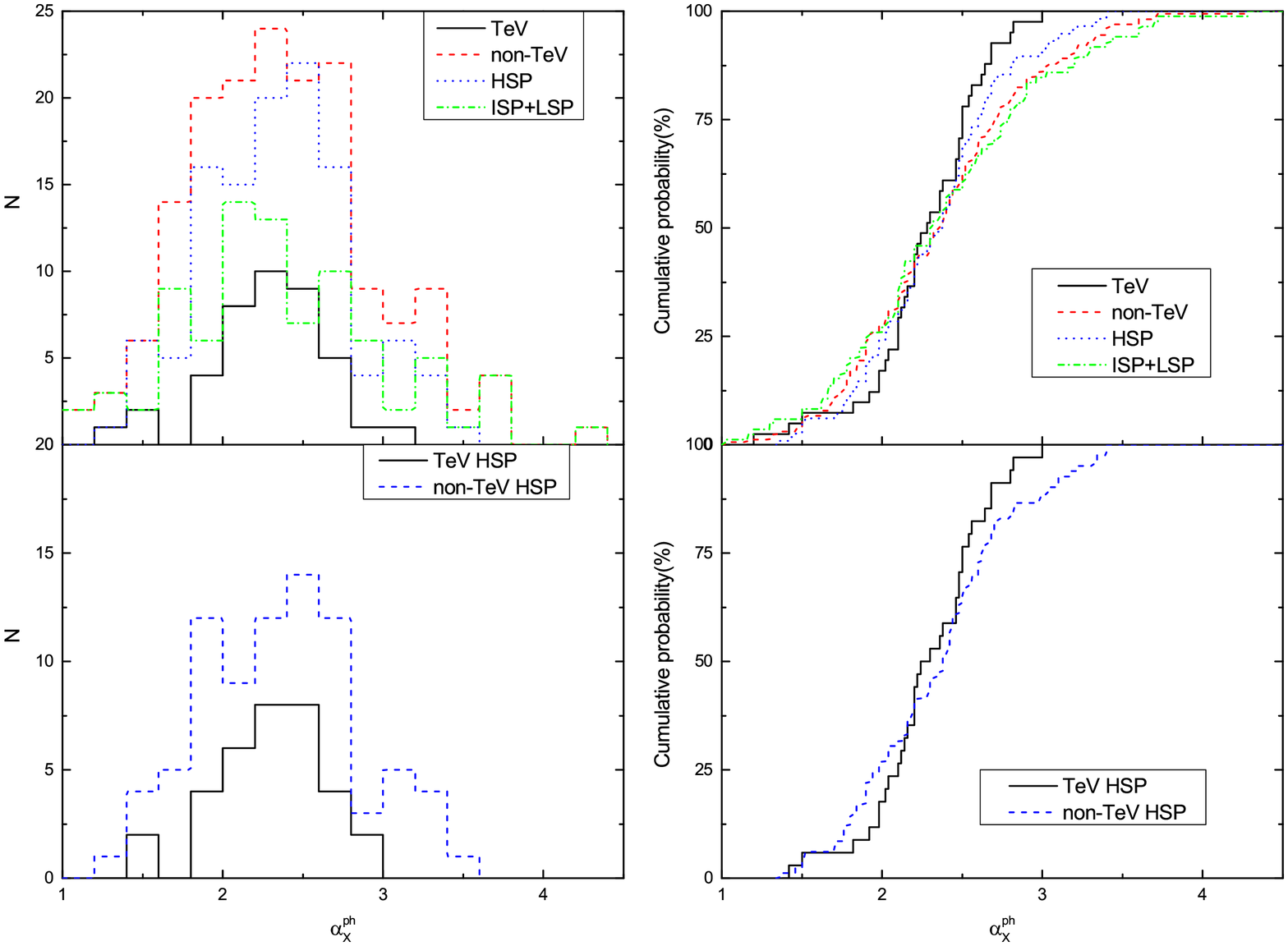}
 \caption{The distribution of the X-ray photon index (left panel) and the accumulative probability (right panel) for the whole sample (upper sub-panel), and for the HSP BL Lacs (lower sub-panel). In the upper sub-panel, both left and right panel, solid line stands for TeV sources, broken line for non-TeV sources, dotted line for HSP, broken-dotted line for ISP+LSP. In the lower sub-panel, both left and right panel, solid line stands for TeV HSP BL Lacs, broken line for non-TeV HSP BL Lacs.}
 \label{Lin-2015-TeV-alphaX-K-S}
 \end{figure}

 \begin{figure}[h]
\centering
 \includegraphics[height=4.50in]{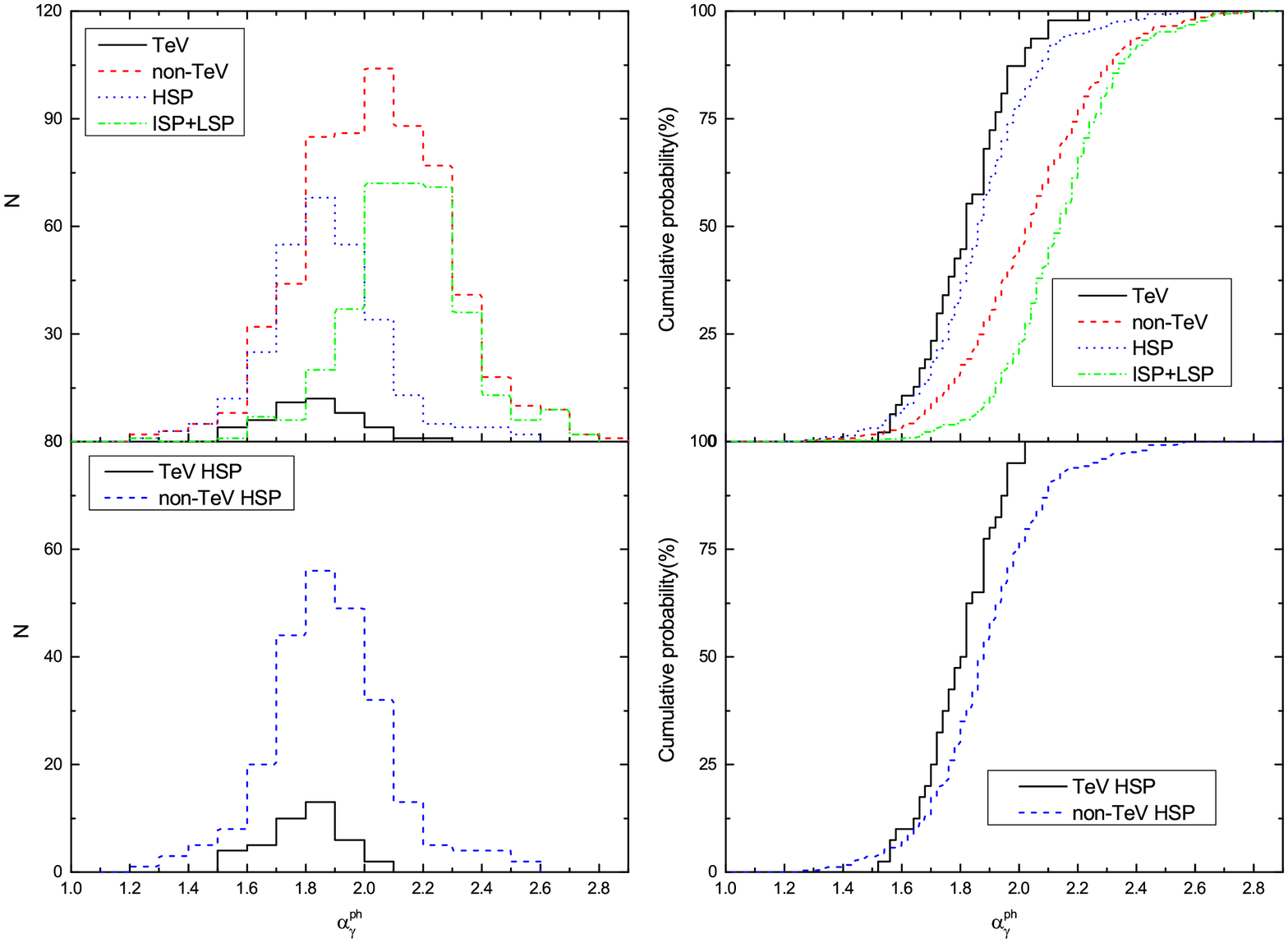}
 \caption{The distribution of the $\gamma$-ray photon index (left panel) and the accumulative probability (right panel) for the whole sample (upper sub-panel), and for the HSP BL Lacs (lower sub-panel). In the upper sub-panel, both left and right panel, solid line stands for TeV sources, broken line for non-TeV sources, dotted line for HSP, broken-dotted line for ISP+LSP. In the lower sub-panel, both left and right panel, solid line stands for TeV HSP BL Lacs, broken line for non-TeV HSP BL Lacs.}
 \label{Lin-2015-TeV-alpha_G-K-S}
 \end{figure}

 \begin{figure}[h]
\centering
 \includegraphics[height=4.50in]{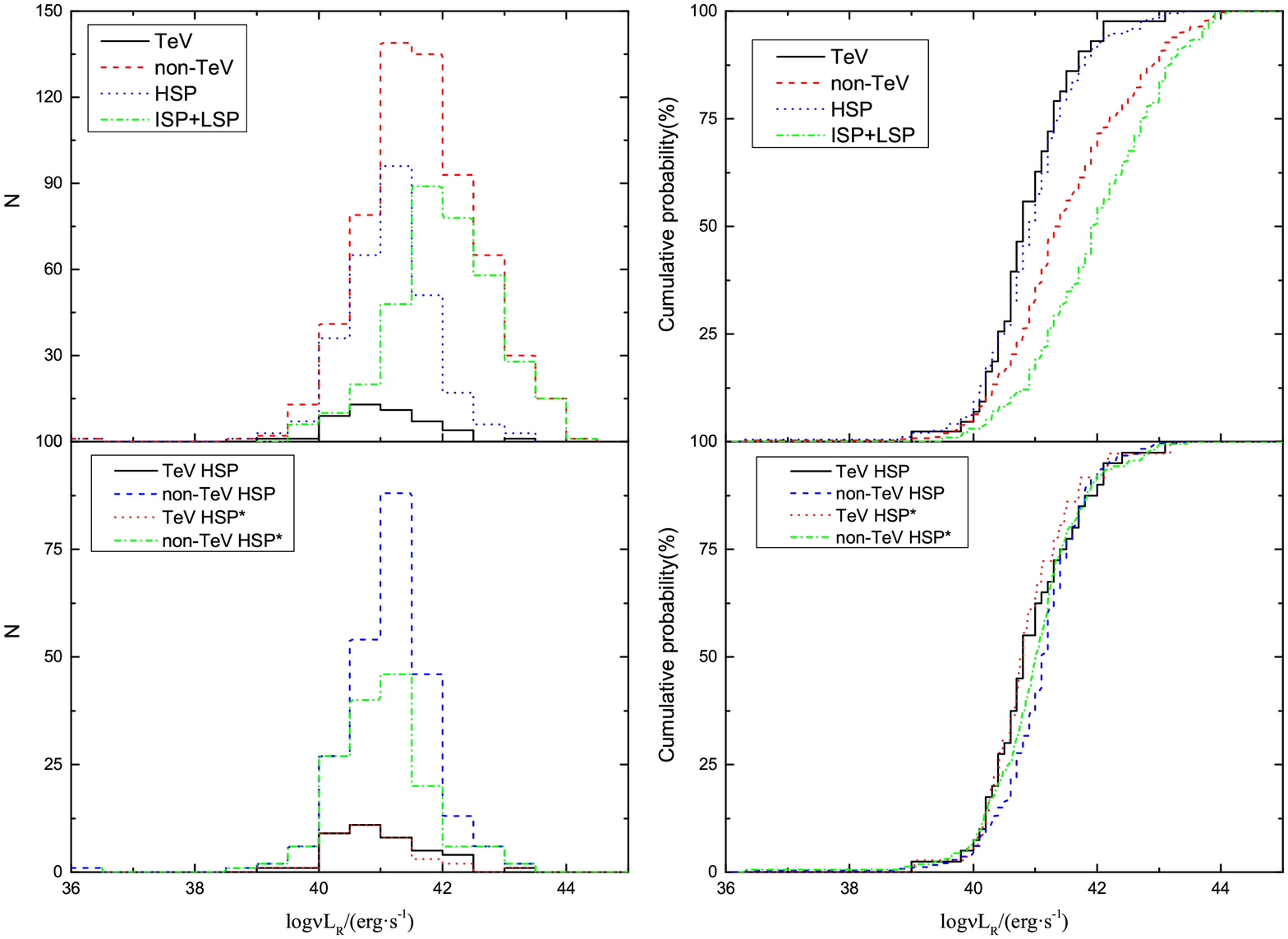}
 \caption{The distribution of the radio luminosity (left panel) and the accumulative probability (right panel) for the whole sample (upper sub-panel), and for the HSP BL Lacs (lower sub-panel). In the upper sub-panel, both left and right panel, solid line stands for TeV sources, broken line for non-TeV sources, dotted line for HSP, broken-dotted line for ISP+LSP. In the lower sub-panel, both left and right panel, solid line stands for TeV HSP BL Lacs, broken line for non-TeV HSP BL Lacs, dotted line for TeV HSP BL Lacs with redsiht, broken-dotted line for non-TeV HSP BL Lacs with redshift.}
 \label{Lin-2015-TeV-L_R-K-S}
 \end{figure}

 \begin{figure}[h]
 \centering
 \includegraphics[height=4.50in]{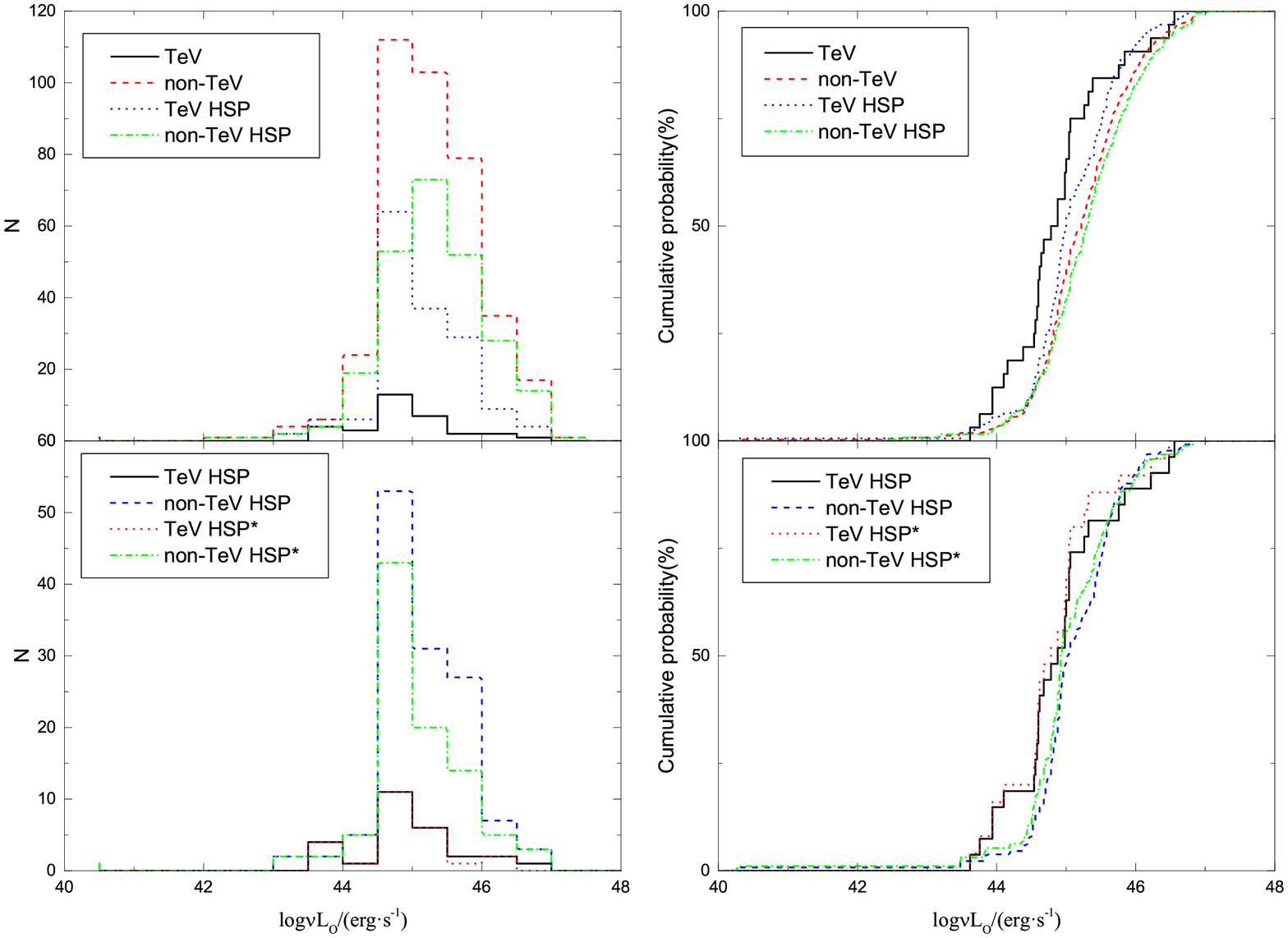}
 \caption{The distribution of the optical luminosity (left panel) and the accumulative probability (right panel) for the whole sample (upper sub-panel), and for the HSP BL Lacs (lower sub-panel). In the upper sub-panel, both left and right panel, solid line stands for TeV sources, broken line for non-TeV sources, dotted line for HSP, broken-dotted line for ISP+LSP. In the lower sub-panel, both left and right panel, solid line stands for TeV HSP BL Lacs, broken line for non-TeV HSP BL Lacs, broken line for non-TeV HSP BL Lacs, dotted line for TeV HSP BL Lacs with redsiht, broken-dotted line for non-TeV HSP BL Lacs with redshift.}
 \label{Lin-2015-TeV-L_O-K-S}
 \end{figure}

 \begin{figure}[h]
 \centering
 \includegraphics[height=4.50in]{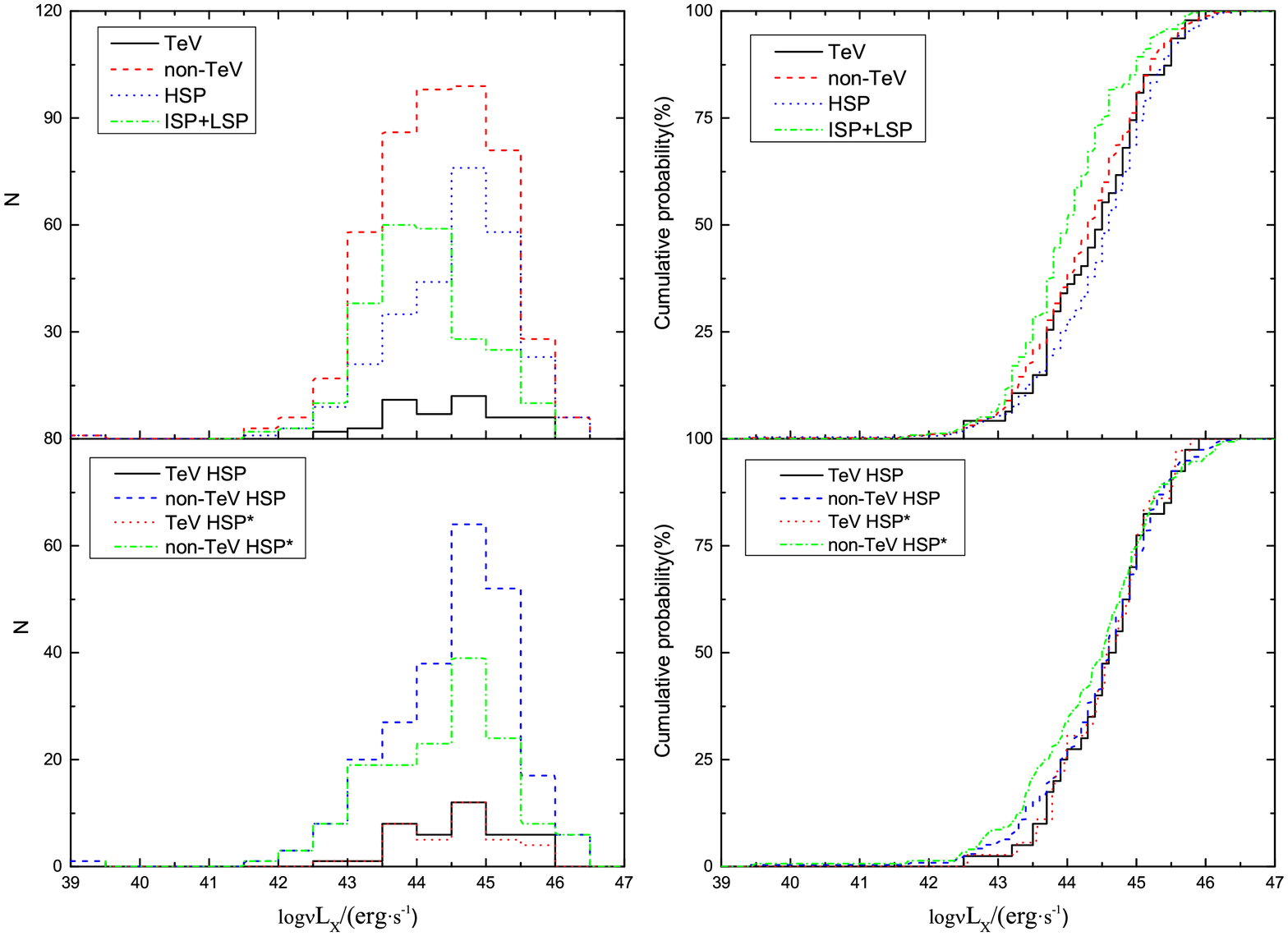}
 \caption{The distribution of the X-ray luminosity (left panel) and the accumulative probability (right panel) for the whole sample (upper sub-panel), and for the HSP BL Lacs (lower sub-panel). In the upper sub-panel, both left and right panel, solid line stands for TeV sources, broken line for non-TeV sources, dotted line for HSP, broken-dotted line for ISP+LSP. In the lower sub-panel, both left and right panel, solid line stands for TeV HSP BL Lacs, broken line for non-TeV HSP BL Lacs, broken line for non-TeV HSP BL Lacs, dotted line for TeV HSP BL Lacs with redsiht, broken-dotted line for non-TeV HSP BL Lacs with redshift.}
 \label{Lin-2015-TeV-L_X-K-S}
 \end{figure}

 \begin{figure}[h]
 \centering
 \includegraphics[height=4.50in]{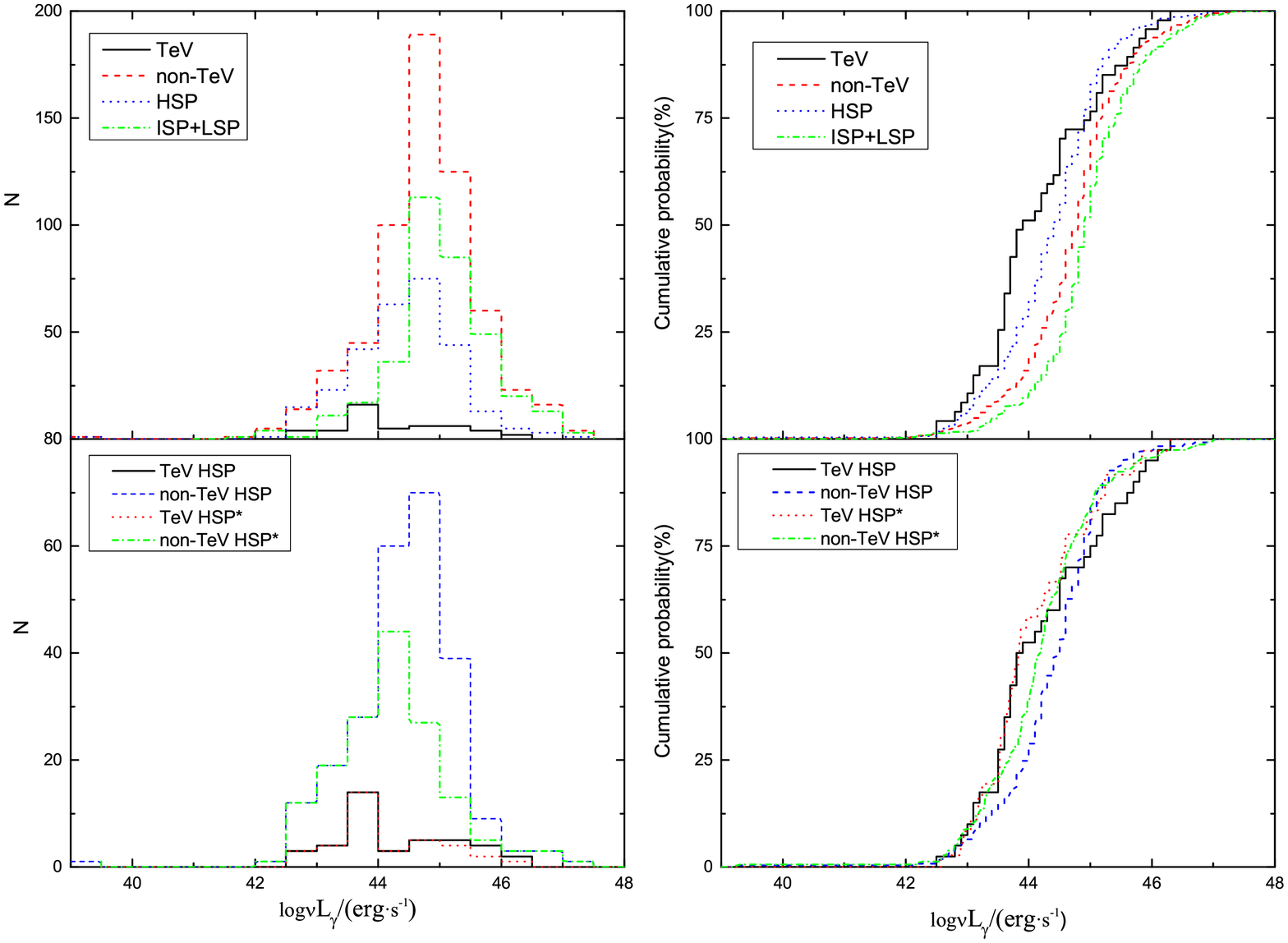}
 \caption{The distribution of the $\gamma$-ray luminosity (left panel) and the accumulative probability (right panel) for the whole sample (upper sub-panel), and for the HSP BL Lacs (lower sub-panel). In the upper sub-panel, both left and right panel, solid line stands for TeV sources, broken line for non-TeV sources, dotted line for HSP, broken-dotted line for ISP+LSP. In the lower sub-panel, both left and right panel, solid line stands for TeV HSP BL Lacs, broken line for non-TeV HSP BL Lacs, broken line for non-TeV HSP BL Lacs, dotted line for TeV HSP BL Lacs with redsiht, broken-dotted line for non-TeV HSP BL Lacs with redshift.}
 \label{Lin-2015-TeV-L_G-K-S}
 \end{figure}

 \begin{figure}[h]
 \centering
 \includegraphics[height=4.50in]{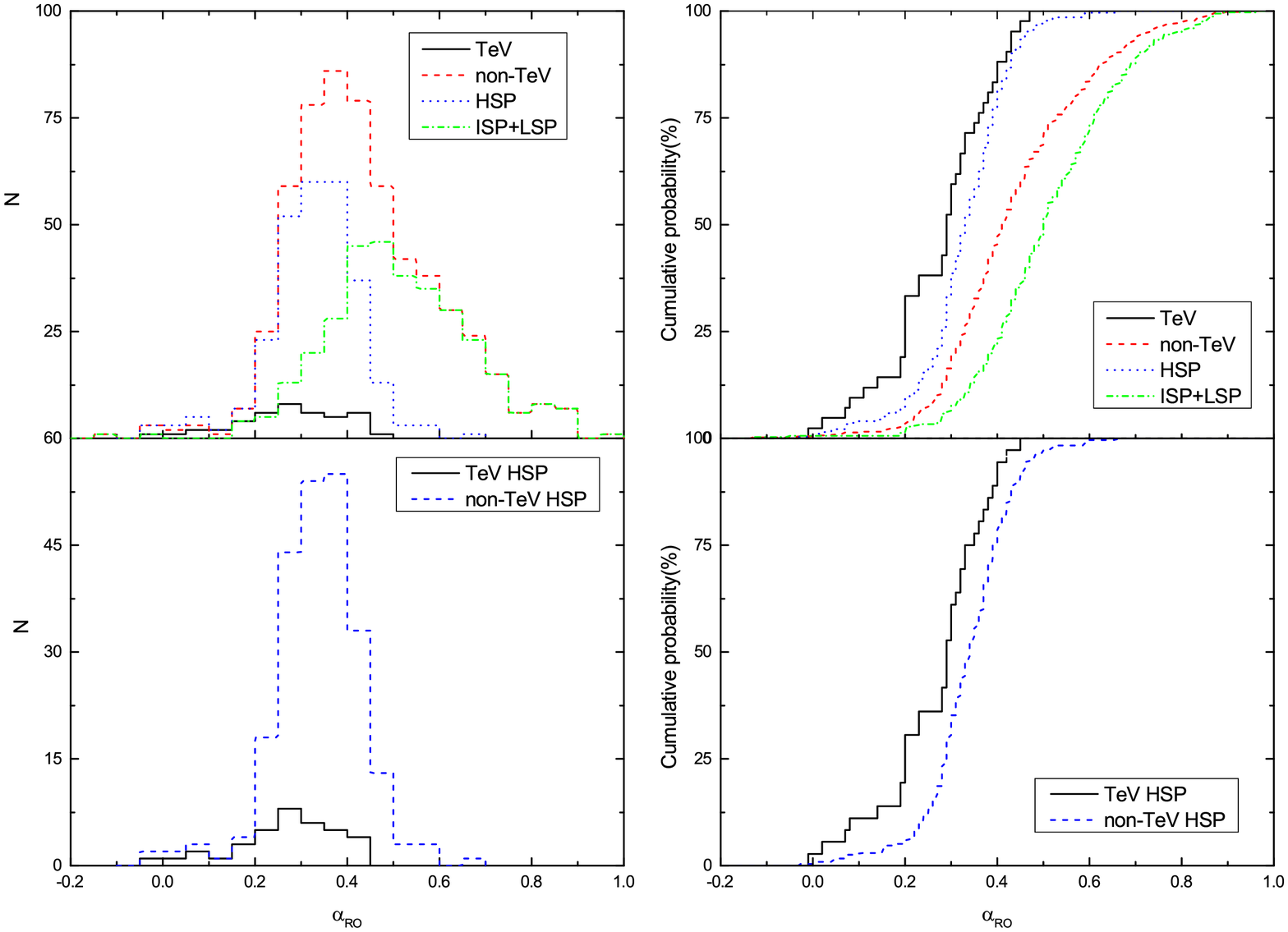}
  \caption{The distribution of the effective spectral index $\alpha_{\rm RO}$ (left panel) and the accumulative probability (right panel) for the whole sample (upper sub-panel), and for the HSP BL Lacs (lower sub-panel). In the upper sub-panel, both left and right panel, solid line stands for TeV sources, broken line for non-TeV sources, dotted line for HSP, broken-dotted line for ISP+LSP. In the lower sub-panel, both left and right panel, solid line stands for TeV HSP BL Lacs, broken line for non-TeV HSP BL Lacs.}
  \label{Lin-2015-TeV-alpha_RO-K-S}
 \end{figure}

 \begin{figure}[h]
 \centering
 \includegraphics[height=4.50in]{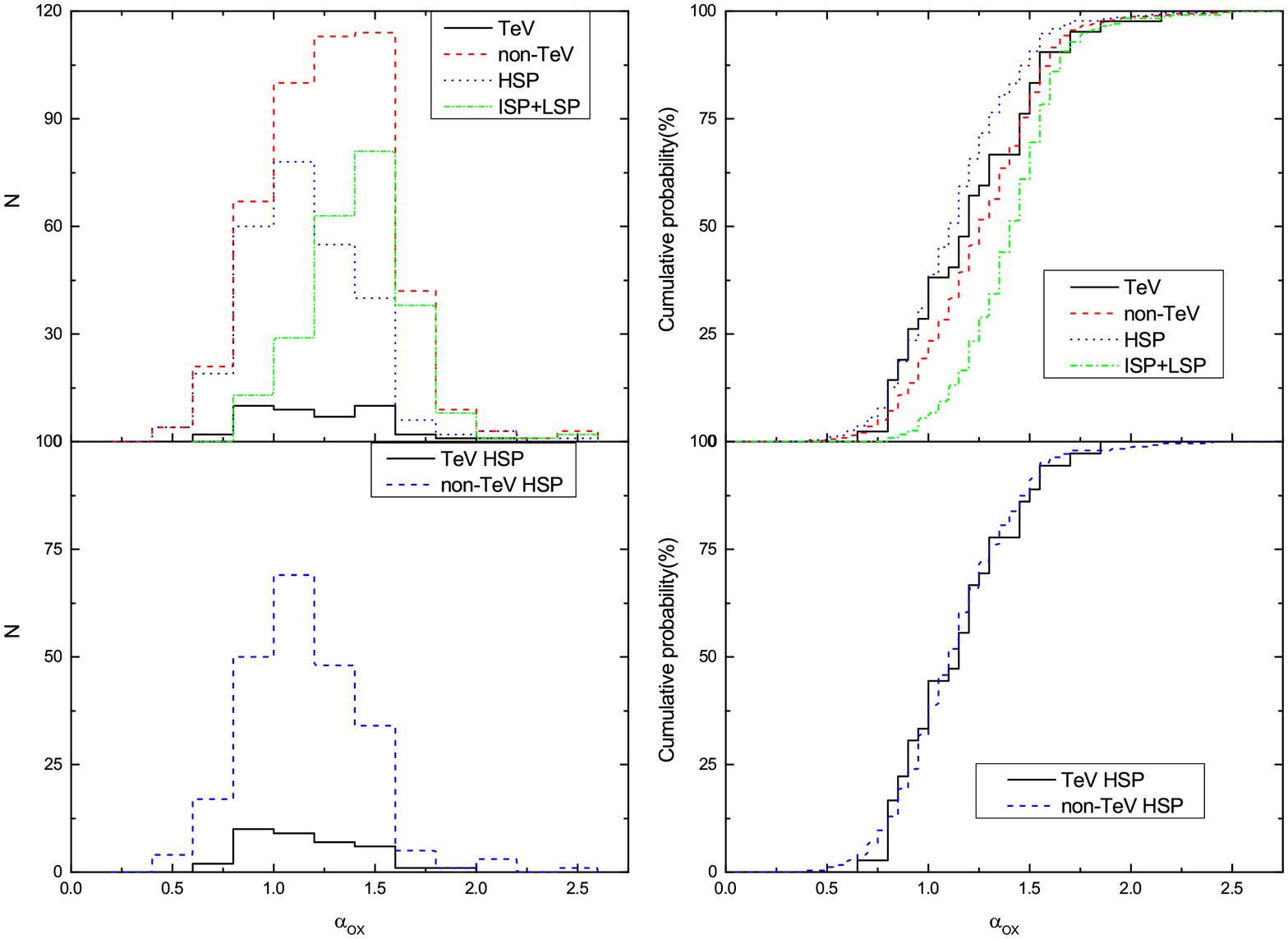}
 \caption{The distribution of the effective spectral index $\alpha_{\rm OX}$ (left panel) and the accumulative probability (right panel) for the whole sample (upper sub-panel), and for the HSP BL Lacs (lower sub-panel). In the upper sub-panel, both left and right panel, solid line stands for TeV sources, broken line for non-TeV sources, dotted line for HSP, broken-dotted line for ISP+LSP. In the lower sub-panel, both left and right panel, solid line stands for TeV HSP BL Lacs, broken line for non-TeV HSP BL Lacs.}
 \label{Lin-2015-TeV-alpha_OX-K-S}
 \end{figure}

\begin{figure}[h]
 \centering
 \includegraphics[height=6.00in]{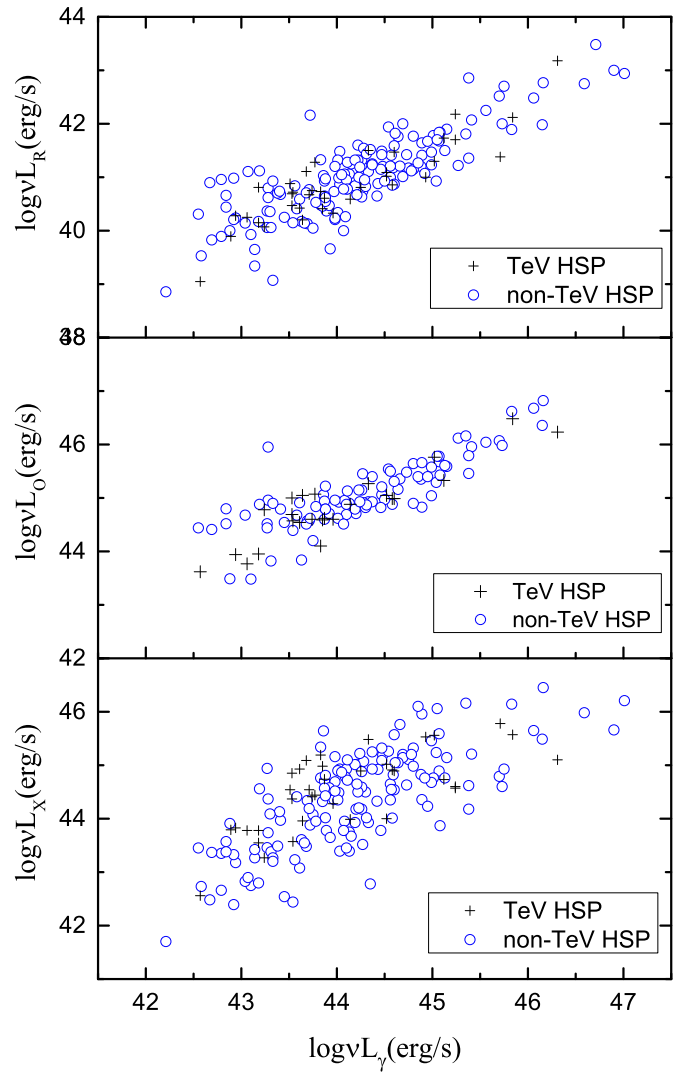}\\
 \caption{Plot of the radio (top), optical (median), X-ray (bottom) luminosities against the $\gamma$-ray luminosity for HSP BL Lacs. In each panel, the cross symbols stand for the TeV HSP BL Lacs, and the circle symbols stand for the non-TeV HSP BL Lacs.}
 \label{Lin-2015-TeV-L-L-corr}
 \end{figure}

 \begin{figure}[h]
 \centering
\includegraphics[height=4.50in]{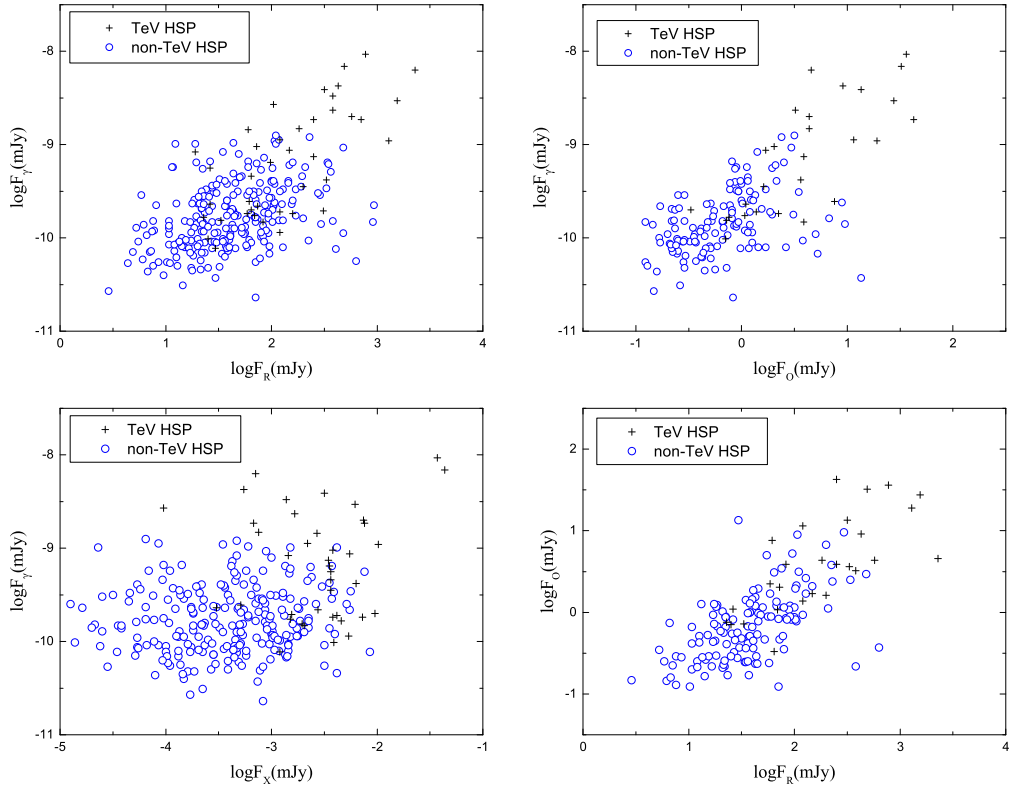}\\
 \caption{Plot of the $\gamma$-ray flux against the radio flux (left upper panel), the optical flux (right upper panel), and the X-ray flux (left lower panel) for HSP BL Lacs. Plot of the optical flux against the radio flux for HSP BL Lacs in right lower panel. In each panel, the cross symbols stand for the TeV HSP BL Lacs, and the circle symbols stand for the non-TeV HSP BL Lacs.}
 \label{Lin-2015-TeV-logF-logF-corr}
 \end{figure}

\begin{figure}[h]
 \centering
 \includegraphics[height=6.00in]{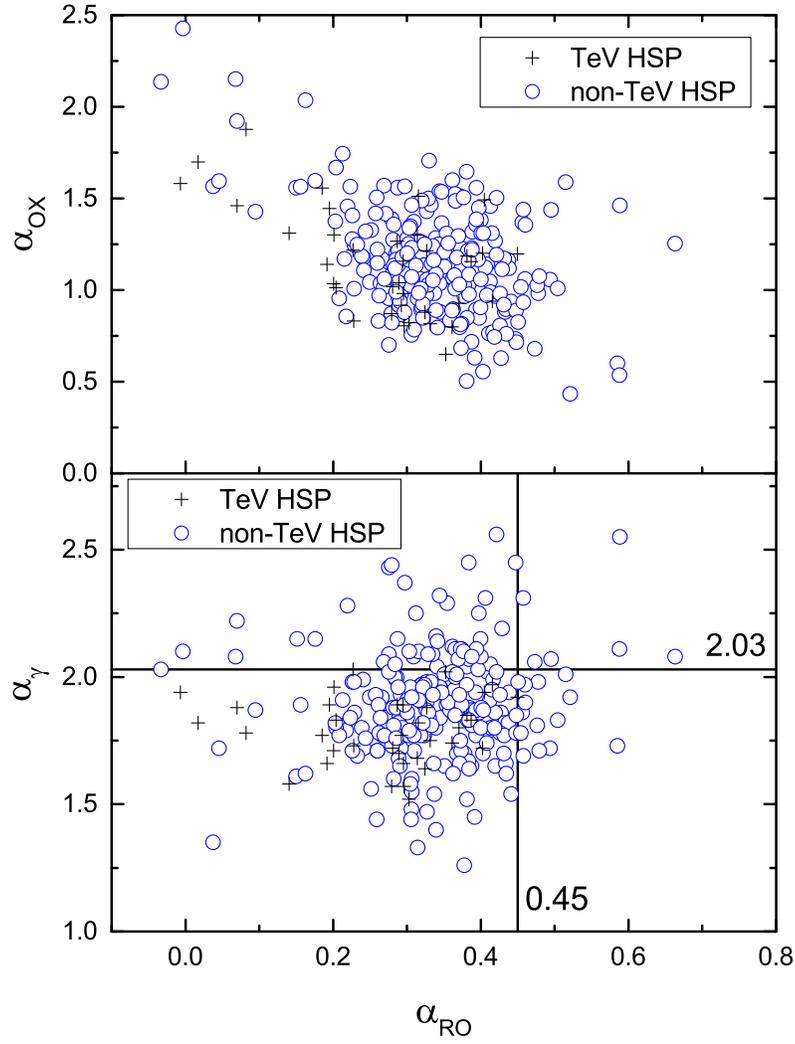}\\
 \caption{Plot of the effective spectral index $\alpha_{\rm OX}$ (upper panel) and $\gamma$-ray photon index (lower panel) against the effective spectral index $\alpha_{\rm RO}$ for HSP BL Lacs. In each panel, the circle symbols stand for the non-TeV HSP BL Lacs, and the cross symbols stand for the TeV HSP BL Lacs.}
 \label{Lin-2015-TeV-alpha_RO-alpha_OX-alpha_G}
 \end{figure}

\end{document}